\documentclass[preprint2]{aastex}

\newcommand{\ivcg}{{G86~}}
\newcommand{\casa}{{Cas A~}}

\slugcomment{To appear in the Astrophysical Journal}

\shorttitle{Map-making method for timestream data from large arrays}
\shortauthors{Patanchon et al.}

\begin{document}

\title{SANEPIC: A Map-Making Method for Timestream Data From Large Arrays}

\author{G.~Patanchon,\altaffilmark{1,2,\dag}
        P.~A.~R.~Ade,\altaffilmark{3}
        J.~J.~Bock,\altaffilmark{4,5}
        E.~L.~Chapin,\altaffilmark{1}
        M.~J.~Devlin,\altaffilmark{6}
        S.~Dicker,\altaffilmark{6}
        M.~Griffin,\altaffilmark{3}
        J.~O.~Gundersen,\altaffilmark{7}
        M.~Halpern,\altaffilmark{1}
        P.~C.~Hargrave,\altaffilmark{3}
        D.~H.~Hughes,\altaffilmark{8}
        J.~Klein,\altaffilmark{6}
        G.~Marsden,\altaffilmark{1}
        P.~G.~Martin,\altaffilmark{9,10}
        P.~Mauskopf,\altaffilmark{3}
        C.~B.~Netterfield,\altaffilmark{10,11}
        L.~Olmi,\altaffilmark{12,13}
        E.~Pascale,\altaffilmark{11}
        M.~Rex,\altaffilmark{6}
        D.~Scott,\altaffilmark{1}
        C.~Semisch,\altaffilmark{6}
        M.~D.~P.~Truch,\altaffilmark{14}
        C.~Tucker,\altaffilmark{3}
        G.~S.~Tucker,\altaffilmark{14}
        M.~P.~Viero,\altaffilmark{10}
        D.~V.~Wiebe\altaffilmark{11}}

\altaffiltext{1}{Department of Physics \& Astronomy, University of
British Columbia, 6224 Agricultural Road, Vancouver, BC V6T~1Z1,
Canada}

\altaffiltext{2}{Laboratoire APC, 10, rue Alice Domon et L{\'e}onie Duquet 75205, Paris, France}

\altaffiltext{3}{Department of Physics \& Astronomy, Cardiff University, 5 The Parade, Cardiff, CF24~3AA, UK}

\altaffiltext{4}{Jet Propulsion Laboratory, Pasadena, CA 91109-8099}

\altaffiltext{5}{Observational Cosmology, MS 59-33, California Institute of Technology, Pasadena, CA 91125}

\altaffiltext{6}{Department of Physics and Astronomy, University of Pennsylvania, 209 South 33rd Street, Philadelphia, PA 19104}

\altaffiltext{7}{Department of Physics, University of Miami, 1320 Campo Sano Drive, Carol Gables, FL 33146}

\altaffiltext{8}{Instituto Nacional de Astrof\'isica \'Optica y Electr\'onica (INAOE), Aptdo. Postal 51 y 72000 Puebla, Mexico}

\altaffiltext{9}{Canadian Institute for Theoretical Astrophysics, University of Toronto, 60 St. George Street, Toronto, ON M5S~3H8, Canada}

\altaffiltext{10}{Department of Astronomy \& Astrophysics, University of Toronto, 50 St. George Street, Toronto, ON  M5S~3H4, Canada}

\altaffiltext{11}{Department of Physics, University of Toronto, 60 St. George Street, Toronto, ON M5S~1A7, Canada}

\altaffiltext{12}{Istituto di Radioastronomia, Largo E. Fermi 5, I-50125, Firenze, Italy}

\altaffiltext{13}{University of Puerto Rico, Rio Piedras Campus, Physics Dept., Box 23343, UPR station, San Juan, Puerto Rico}

\altaffiltext{14}{Department of Physics, Brown University, 182 Hope Street, Providence, RI 02912}

\altaffiltext{\dag}{\url{patanchon@apc.univ-paris7.fr}}

\begin{abstract}
  We describe a map-making method which we have developed for the
  Balloon-borne Large Aperture Submillimeter Telescope (BLAST)
  experiment, but which should have general application to data from
  other submillimeter arrays.  Our method uses a Maximum Likelihood
  based approach, with several approximations, which allows images to
  be constructed using large amounts of data with fairly modest
  computer memory and processing requirements.  This new approach,
  Signal And Noise Estimation Procedure Including Correlations
  (SANEPIC), builds upon several previous methods, but focuses
  specifically on the regime where there is a large number of detectors
  sampling the same map of the sky, and explicitly allowing for the 
  the possibility of strong correlations between the detector timestreams.
  We provide real and simulated examples of how well this method performs
  compared with more simplistic map-makers based on filtering.  We discuss
  two separate implementations of SANEPIC: a brute-force approach, in which
  the inverse pixel-pixel covariance matrix is computed;
  and an iterative approach, which is much more efficient for large maps.
  SANEPIC has been successfully used to produce maps using data from the
  2005 BLAST flight.
\end{abstract}


\hyphenation{Sub-milli-meter}
\hyphenation{sub-milli-meter}

\section{Introduction}

The problem of optimal ``map-making'' or ``image reconstruction'' is
complex and multi-faceted, with the basic procedures and even the
terminology differing dramatically between different sub-fields of
astronomy.  The method adopted depends on the form in which the data
are gathered, and on the dominant source of systematic effects.  From
the optical to the near-IR one talks about combining ``frames'', along
with measurements of ``darks'' and ``flat-fields''.  For the reduction
of Cosmic Microwave Background (CMB) data the now conventional method
is to start from the principle that there is a linear algebra approach
to solving the Maximum Likelihood problem.  However, this has only
been feasible up until now because of the limited number of detectors in the
typical CMB experiment, and the fact that correlated signals among the
detectors can be effectively ignored.  Because of the rapid development of
large bolometer arrays, the question which arises is:
how does one adapt the CMB approach to dealing with substantial numbers of
detectors, and where there are significant cross-correlations of
noise between the detector timestreams.

Data from the Balloon-borne Large Aperture Submillimeter Telescope
(BLAST, Devlin et al.~2004) represent a new challenge for bolometric
timestream-to-map algorithms.  Recent CMB experiments which use
detectors similar to those used on BLAST, such as BOOMERANG
\citep{crill} and Archeops \citep{benoit}, only use a handful of
separate bolometers.  Furthermore, these experiments' off-axis designs
lead to small correlations between detectors.  Consequently, the
correlations could be ignored at the map-making stage, and each
detector timestream could be treated as an independent sub-set of the
data.  This has changed for BLAST, which has up to 139 detectors per
band, with significant correlations induced by the on-axis design, as
well as the higher frequencies of the observations.  Just by itself,
the large number of channels increases the impact of even small time
stream correlations, the contribution from which does not integrate
down with increasing number of detectors, unlike the uncorrelated
noise.  The high level of correlation (largely induced by temperature
drifts in the obscuring secondary supports in BLAST) make it important
that the correlations be handled carefully in the map-making process.

This paper describes a Signal And Noise Estimation Procedure Including
Correlations  (SANEPIC) which has been developed for the analysis of
BLAST.  This algorithm will also have application to many next
generation experiments which will involve both noise correlations
between channels (including correlations from the atmosphere) and very
large numbers of detectors.  This includes the next generation of
larger-format arrays for use in ground- and balloon-based instruments
at microwave and millimeter wavelengths, which will have typically
thousands of detectors.

This paper is arranged as follows.  We next describe the pertinent
aspects of BLAST.  In Section~3, the longest section, we set out our
basic map-making method.  Simulations we use for testing the method
are described in Section~4, with results presented in Section~5,
demonstrating the benefit of accounting for the correlated noise in
BLAST-like observations.  Finally some of the maps obtained from the
June 2005 BLAST flight are presented in Section~6.

\section{BLAST observations of the submillimeter sky}

The map-making procedure presented in this paper has been used to
analyze the data from the BLAST 2005 flight. We will use BLAST as a
specific example for the application of SANEPIC throughout the paper.
We describe BLAST in Section~\ref{sub:BLAST}, we then summarize the
pre-processing of the data prior to map-making in Section
\ref{sub:process}. In Section~\ref{sub:model}, we derive a model of
the data that we will use for the map-making.

\subsection{BLAST instrument and observations}
\label{sub:BLAST}

The Balloon-borne Large-Aperture Submillimeter Telescope incorporates
a 2-m primary mirror, and large format bolometer
arrays operating at 250, 350, and 500$\,\mu$m, designed to have 144, 96
and 48 bolometers, respectively (of which 139, 88 and 43,
respectively, were used).  The instrument is described in detail in
\cite{pascale07}.  The low atmospheric opacity at the operating
altitude of ${\sim}\,38\,$km allows BLAST to map the sky very quickly
compared to ground-based experiments and to conduct large area shallow
surveys as well as very deep surveys of the sky \citep{devlin}.  The
BLAST wavelengths are near the peak of the spectral energy
distribution of cold galactic dust, which gives BLAST the ability to
conduct unique extragalactic and Galactic submillimeter surveys with
high spatial resolution and sensitivity.  BLAST thus enables studies
of the distribution of very high-redshift galaxies and of star forming
regions in our Galaxy.

The typical observing strategy consists of scanning the telescope back
and forth in azimuth, covering the entire field by slowly varying the
elevation. Cross-linking of the data is assured by scanning the same
field at another time of the day. Typical scanning strategies are
given in \cite{pascale07}.

The first scientific flight of BLAST took place in June 2005 from the
Esrange Arctic base in Sweden to the Canadian Arctic. A total of
$\sim$ 100 hours of data were taken in a variety of Galactic
fields. They include a star forming region (Vulpecula) over $4\,{\rm deg}^2$,
described in \cite{chapin}, three other fields of similar size
in the Galactic Plane (which will be the focus of future papers), an
integration towards the ELAIS-N1 field (see Oliver et al.~2000),
the \casa supernovae remnant over about $0.5\,{\rm deg}^2$
(Hargrave et al.~in preparation),
and several compact Galactic and extra-galactic sources \citep{truch07}.
Hereafter we refer to these as the BLAST05 data, to distinguish them from
the data taken during the December 2006 Antarctic flight.

\subsection{Time-ordered data pre-processing}
\label{sub:process}

The processing of BLAST data from detector timestreams to the final
map product involves several steps prior to map-making. Each of these
steps is designed to remove a particular (or several) artifact(s) from
the data, and sometimes requires iterating, since some effects need to
be removed simultaneously. In the following, we summarize the main
processing stages leading to the time-ordered segments which are used
as inputs for the map-making process.

We start by identifying events in the data which are sharply localized
in time, such as spikes from cosmic rays hits and other spurious
sources. We use a method which allows us to discriminate between the
different events depending on their signature in the data.  Spikes
which involve only a single sample are flagged and the corrupted
samples are replaced by the average value of the samples in the
vicinity. The data are deconvolved from the low-pass filter applied by
the readout electronics. This filter has a frequency cut off of
approximately $35\,$Hz and is designed to avoid high frequency noise
aliasing. The deconvolution is performed in Fourier space. In
addition, we have applied a low-pass cosine filter which limits the
noise power from increasing at very high frequency (above $38\,$Hz) due to
deconvolution. We have checked that the noise power spectrum is
relatively flat after these deconvolution operations.
Finally, cosmic ray hits
and other localized artifacts in the data timestreams are detected and cut
out.  In order to avoid biasing our data products by having
systematically more false event detections located where the sky is
bright (e.g.,~when scanning a point source), we iterate this process
-- we make maps starting from data which has been cleaned using the
process described above, subtract the maps from each original data-set,
and reprocess the data. The maps calculated at this intermediate
stage are obtained by simply rebinning the data into pixels after
strongly high-pass filtering. The filter applied is a Butterworth
filter with a frequency cut off of $0.5\,$Hz, which is of the order of
the knee frequency of the noise. Even if this operation suppresses
most of the intermediate to large scales of the sky signal in the
maps, it does not very much alter the signal from localized sources, at
least to the level of accuracy needed at this stage, and it has the
advantage of removing most of the stripes in the maps due to $1/f$
noise. We have verified that the resulting bias for bright calibrators
is less than 1\%.

About 2\% of the data from the BLAST05 flight was removed due to cosmic
ray events. Most of the events affected a single detector timestream,
although some events affected a whole array at the same time. Detected
spikes (from cosmic rays but also other spurious effects) are flagged
over small time intervals of typically 1~second in the data. One second is
too large an interval to simply ignore, so the corrupted data need to be
replaced with random noise generated in a way that as much as possible
preserves the statistical properties of the data. In the later
map-making stage, which is partly performed in Fourier space, we
assume continuity of the data.  We could perform this gap-filling with
a constrained noise realization (see e.g., Stompor et al. 2002).  However,
since the gap intervals are significantly smaller than the inverse of the knee
frequency of the noise power spectrum ($0.3\,$Hz), the noise can be well
approximated by the sum of a white component plus a straight line of
some slope across the gap.  Specifically, we generate white noise in each gap
with a standard deviation measured from the data in the vicinity of
the gap, and add a baseline with the parameters fit using 20
samples on each side. The white noise generation is for restoring as
best as possible the stationarity of the data (generated samples are
not reprojected to the map at the end).

After having filled the gaps in the timestreams, we filter out very low
frequency drifts which are poorly accounted for in the map-making procedure. A
fifth-order polynomial is fit to the data and removed from each data
segment in order to reduce fluctuations on timescales larger than the
length of the considered segment which, depending on the specific
case, varies from 30 minutes to a few hours. These fluctuations are
poorly described because of the limited number of Fourier modes, and
would cause leakage at all timescales (for instance a gradient in the
timestream is described by a wide range of Fourier modes), degrading
the efficiency of map-making.  Note that we experimented with various
polynomials and other effective high-pass filters; we found that the
results were not very sensitive to precisely how this is done, but
some such filtering is certainly required. The degree of the polynomial
was chosen empirically as a compromise between suppressing the artifacts
and keeping the large scale signals in the final maps.
Using simulations,
we have checked that the effect on the transfer function of the signal
in the final map is weak. The resulting data segments are then corrected
for the time-varying calibration (see Truch et al.~2007) using
measurements of an internal calibration lamp \citep{pascale07}.
Finally, the data segments are apodized at the edges over ${\sim}\,2,000$
samples and are high-pass filtered at $5\times 10^{-3}\,$Hz with a
Butterworth filter. This filtering has very little effect on the final
maps, since modes in the data at lower frequencies that are cut in
this way are not expected to contribute significantly to the signals.  We
discuss the choices of filters further in Section~\ref{sub:implem}.

Accurate pointing reconstruction is a complicated procedure for
balloon-borne telescopes, and this affects the map-making task through the
pointing matrix (defined in Section~\ref{sub:model}).
The pointing reconstruction procedure is
described in detail in \cite{pascale07}.  The next important step in
reducing the BLAST data, which is calibration of the detectors, is detailed in
\cite{truch07}.

\subsection{Model of the data}
\label{sub:model}

Having performed the cleaning procedure described in the previous
sub-section, the resulting data timestreams can be modeled very
accurately as the sum of pure signal and pure noise contributions.
The data for detector $i$ observing at a given wavelength and at time
sample $t$ can be written as
\begin{equation}
  d_{it} = [A_i]_{tp} ~s_{p} + n_{it},
  \label{eq:linmodel}
\end{equation}
where $p$ labels the pixels in the final map, $A_i$ is the pointing
matrix for bolometer $i$ (whose elements, indexed with time $t$ and
pixel $p$, give the weight of the contribution of pixel $p$ to the
sample at time $t$ for bolometer $i$), $s_{p}$ is the signal amplitude
at pixel $p$, and the noise amplitude at time $t$ for bolometer $i$ is
$n_{it}$. Summation over repeated indices is assumed here.

Ideally, the element $[A_i]_{tp}$ of the pointing matrix is equal to
the value of the beam response $b( R (\vec{r}-\vec{r_0}))$, where
$\vec{r}$ points to the pixel $p$ location, $\vec{r_0}$ is the
location of the beam center at time $t$, and $R$ is a rotation matrix
which depends on the rotation angle at time $t$ between the telescope
and sky coordinate systems.  In principle one could then recover a map
of the sky {\it deconvolved\/} with the instrumental Point Spread
Function (PSF).  However, in practice this would be unacceptably
noisy, as well as computationally intractable, because of the
prohibitive volume of data.  Even although $A_i$ might have mostly
zero elements, it is nevertheless a huge matrix.  It {\it may\/} be
feasible to deconvolve a non-trivial beam response (e.g.~like the
BLAST05 beams, as shown in Truch et al.~2007) at the same time
performing the map-making step, through an approximate treatment of
the sparse pointing matrix.  But we did not pursue that approach here.

In the simple case where the beam is
symmetric, the map-making problem becomes tractable, provided one restricts
oneself to reconstructing a map of the instrument-convolved sky.
We can then consider $s_p$ in
Equation~\ref{eq:linmodel} as the map of the sky convolved with the
beam, and consequently $A_i$ indicates simply where the detector points in
the sky at a given time. In this case, $A_i$ is an ultimately sparse
matrix with, in the BLAST case, simply a 1 in a single entry of each
row.  This approach, which has been conventional for CMB map-making (although
with some adaptation for chopped data), is what we use in the following
analysis.  It gives no loss of information provided that the map pixels are
sufficiently small, and one can simply
assign all the flux from a bolometer's to the map pixel to
which it points at each time interval.  The
requirement for accuracy is that the pixel size is
smaller by a factor three or more than the Full Width at Half Maximum
(FWHM) of the instrumental PSF; in that case the additional convolution with
the pixel shape gives negligible loss of angular resolution.

The noise term $n_{it}$ in the model represents the sum of all
contributions to the timestreams which do not reproject on the sky.
This will in general include
instrumental noise, fluctuations in atmospheric emission
and other loading and unrecognized cosmic ray hits.  In general, some
of those noise contributions will induce strong correlations
between detector timestreams. In this paper, we adopt a very general
model of the noise where the noise covariance matrix:
\begin{equation}
  N_{ii'tt'} = \left\langle n_{it}\cdot n_{i't'}^t\right\rangle,
\label{eq:modfullN}
\end{equation}
(for bolometer indices $i$, $i'$ and time indices $t$, $t'$) has possibly
non-zero elements even for $i \neq i'$. A key assumption, as we will
see later, is that the noise is Gaussian (so that $N_{ii'tt'}$ is
sufficient to describe all the statistical properties of the noise)
and stationary (constraining $N_{ii'tt'}$, see Sections~\ref{sub:NN}
and \ref{sub:detdetcorr}).

In the specific case of BLAST05 observations, a very significant
correlation of the noise is found in the timestreams, and we have
shown that a more constraining model provides a very good description
of the data. An independent component analysis \citep{delabrouille03}
of the data enabled us to find that the noise and its correlations can
be described to a high degree of accuracy by a noise component which
is {\it not\/} correlated between detectors, together with a single
{\it common-mode\/} component seen by all the detectors at a given
wavelength (some correlations is also seen between detectors from
different wavelength but we have chosen to treat each wavelength
independently). The common part of the noise is instantaneous, meaning
that the same common-mode noise is seen at the same time by all the
detectors. In our model, the noise term in Equation~\ref{eq:linmodel}
is then decomposed as:
\begin{equation}
  n_{it} = \tilde{n}_{it} + \alpha_i c_{t},
  \label{eq:lincmod}
\end{equation}
where the first term is the noise which is uncorrelated between
detectors and the second term represents the common-mode component of
the noise, rescaled by an amplitude parameter $\alpha_i$ which depends
on the detector but not on time. This model can be generalized easily
to deal with multiple noise components in timestreams. 

In the following section, we present a method to reconstruct $s_p$
given the data, in the framework of the linear model
(Equation~\ref{eq:linmodel})
and in the presence of correlated noise between detector timestreams.

\section{Map-making method}

\subsection{Maximum Likelihood map-making}

The use of Maximum Likelihood map-making techniques has been developed
by many authors for application to Cosmic Microwave Background data-sets
\citep{wright,tegmark,borrill,prunet00,tegmark00,ferreira,dore,natoli,prunet01,dupac02,stompor,hinshaw,yvon}. Some
other approach are more specific to destriping for {\sl Planck}-like
scanning strategies
\citep{delabrouille98,maino02,keihanen,degasperis,macias,poutanen,ashdown07}.
Since there is already a large number of publications on the topic, here we
present only a very brief overview of the approach of Maximum
Likelihood map-making techniques.

Assuming the simple linear model given by
Equation~\ref{eq:linmodel}, the log-likelihood of the data can be
calculated under the assumption that the noise is Gaussian and
stationary.  The solution is
\begin{equation}
  \log L(d|s) = - \frac12 (d - As)^t N^{-1} (d - As),
\end{equation}
where $N \equiv \left\langle n.n^t\right\rangle$ is the noise
covariance matrix in the time domain, and $.^t$ denotes transpose.
Maximizing the above equation with respect to the map parameters $s$
(suppressing the pixel indices here for convenience) leads to the
following well known estimator:
\begin{equation}
  \hat{s} = (A^tN^{-1}A)^{-1} A^tN^{-1} d.
  \label{eq:solmap}
\end{equation}
The inverse pixel-pixel covariance matrix of the noise in the map is the
term in brackets in this equation, i.e.
\begin{equation}
  N_{pp'}^{-1} = A^tN^{-1}A.
  \label{eq:InvNpix}
\end{equation}
Computation of the solution to Equation~\ref{eq:solmap} is far from
trivial for most astronomical applications, due to the large amount of
data, and hence this poses a difficult numerical challenge.  The noise
covariance matrix $N$ is a very large matrix of size the number of
samples squared, which could easily be millions, while $N_{pp'}$ may be
more reasonable in size but has no obvious symmetries, and so is still
difficult to invert. Nevertheless, we have implemented a method aimed
at finding the Maximum Likelihood solution given by
Equation~\ref{eq:solmap} when there are a large number of detectors
and in the presence of possible correlations in the noise between
different detector timestreams.

\subsection{Implementation}
\label{sub:implem}

In the simple case of dealing only with independent noise between
detectors, our matrix-inversion method is very similar to the MADCAP
method, described in \cite{stompor}.  However, we have developed our new
approach to deal efficiently with multi-detector data in the presence
of correlated noise between detectors (described in detail in
Section~\ref{sub:detdetcorr}).  In this section, we summarize the
basic ideas for the simpler 1-detector 1-scan case.

In order to find the Maximum Likelihood solution of the map
(Equation~\ref{eq:solmap}), we have developed two different
algorithms. They both allow us to solve the linear system
$N_{pp'} {\hat s} = x$, with $x \equiv A^tN^{-1} d$. The first approach
explicitly computes the inverse pixel-pixel covariance matrix $N_{pp'}^{-1}$,
and we refer to this as the `brute-force algorithm'.  The second approach
uses iterations which converge to the Maximum Likelihood map without
the need for computing $N_{pp'}^{-1}$, and we refer to this as the `iterative
algorithm'.  Both approaches require as a first step
the computation of the inverse of the time-time noise covariance matrix $N$.

\subsubsection{Inverse noise covariance matrix $N_{tt'}^{-1}$}
\label{sub:NN}

In practice, even when we have knowledge of the statistical properties
of the data as described by the power spectrum $P(\omega)$, the
brute-force inversion of $N$ is not tractable because of its enormous
size -- for a single BLAST detector observing for 10 hours at a
data-rate of $100\,$Hz, the matrix has approximately $10^{13}$
elements.  However, if we make the approximation that each data
segment is ``circulant'', meaning that the beginning and the end of a
segment are connected without discontinuity and that there are no gaps
in the data, then the matrix $N$ is also circulant (see
Section~\ref{sub:gaps} for a description of how we treat gaps in the
data).  Circulant matrices are much easier to store and to invert.
With this approximation the matrix can be written
\begin{equation}
  N_{tt'} = C(|t-t'|),
\end{equation}
where the correlation function $C(|t-t'|)$ between samples $t$ and
$t'$ depends only on the separation between the two samples.  A
circulant matrix has the property of having a diagonal matrix
counterpart in Fourier space.

Let $F$ be the discrete Fourier operator, we have
\begin{equation}
  N = F^\dagger \Lambda F,
\end{equation}
where $.^\dagger$ denotes transpose conjugation, and $\Lambda$ is a
diagonal matrix whose diagonal is described by the power spectrum of
the data segment,
\begin{equation}
  \Lambda_{\omega\omega} = P(\omega).
\end{equation}
The inverse of the noise covariance matrix is
\begin{equation}
  N^{-1} = F^\dagger \Lambda^{-1} F,
\end{equation}
and because $\Lambda^{-1}$ is a diagonal matrix, $N^{-1}$ is also
circulant, so that knowledge of only one row is enough to describe the
entire matrix.  Then the inverse covariance matrix can be written
\begin{equation}
  [N^{-1}]_{tt'} = \bar{C}(|t-t'|),
\end{equation}
with
\begin{equation}
  \bar{C}(\Delta t) = {\cal{F}}^{-1}\left\{{1 \over P(\omega)}\right\}(\Delta t),
  \label{eq:PktobC}
\end{equation}
where ${\cal{F}}^{-1}$ represents the inverse Fourier transform.  The
inverse of the covariance matrix can then be computed directly using
the power spectrum of the data.  This is a very fast operation
($O(n_{\rm s}\log n_{\rm s})$, with $n_{\rm s}$ the number of samples),
and does not require large memory since only one row of the matrix
is computed.

The approximation that the data segments are ring-shaped or circulant
might seem unreasonable, but in the end this only has an effect on a
small fraction of the matrix.  For data with large-scale correlations
(data described by a $1/f$ power spectrum, for instance), the
approximation implies an assumption that the two edges of the segment are
very correlated.  In reality, there is obviously little correlation at
long timescales compared to short timescales, so extra striping
could be introduced in the maps if one steps across the two edges of the data
segment.  We have addressed this problem in two ways in order to
avoid introducing artifacts in the final map.  In the case where we explicitly
compute the inverse pixel-pixel covariance matrix $N_{pp'}^{-1}$ (as in
the brute-force inversion algorithm), for the estimation of $N$ entering
in the computation of $N^{-1}_{pp'}$, we constrain $\bar{C}(\Delta t) =
0$ for $\Delta t > n_{\rm s}/2$, with $n_{\rm s}$ being the number of
samples in the segment; this is sometimes known as ``the MADCAP
approximation'' in the literature.  It is important to note that
this approximation cannot be used for
the computation of $A^tN^{-1}d$, because it is performed partly in
Fourier space (see Section~~\ref{subsub:ANd}).  Instead we have apodized
the data $d$ at the edges, and we have removed a low-order polynomial
(5th order in practice) to reduce fluctuations having typical
timescales of order (or larger than) the data segment (see Section
\ref{sub:process}).  This is reasonable, since those scales are not
well described by a limited number of Fourier modes, and hence will
always be hard to reconstruct.

\subsubsection{Inverse pixel-pixel covariance matrix $N_{pp'}^{-1}$}
\label{subsub:InvNpp}

The computation of the inverse pixel-pixel covariance matrix, which is
described in this section, is required only in the brute-force inversion
algorithm, or for an accurate error estimation in the pixel domain.

Since the pointing matrix has only one non-zero element per row in our
simple model (one data sample is associated with a single map pixel),
the matrix multiplication $A^tN^{-1}A$ requires a single loop going
across all the non-zero elements of $N^{-1}$.  For most cases,
a dominant faction of the map-making computing time will be devoted to this
operation.  If the data are only correlated within a typical length
$\lambda_{\rm c}$, we have the property: $\bar{C}(\Delta t)\,{\simeq}\,0$
for $\Delta t\,{>}\,\lambda_{\rm c}$, and $N^{-1}$ is a band-diagonal
matrix (elements separated from the diagonal by more than
$\lambda_{\rm c}$ are negligible).  The number of elements to go
through in the loop is of the order of, but smaller than $2n_{\rm
  s}\lambda_{\rm c}$, which is hopefully much smaller than the size of
the matrix itself.

\begin{figure*}[t!]
  \begin{center}
    \includegraphics[width=\columnwidth]{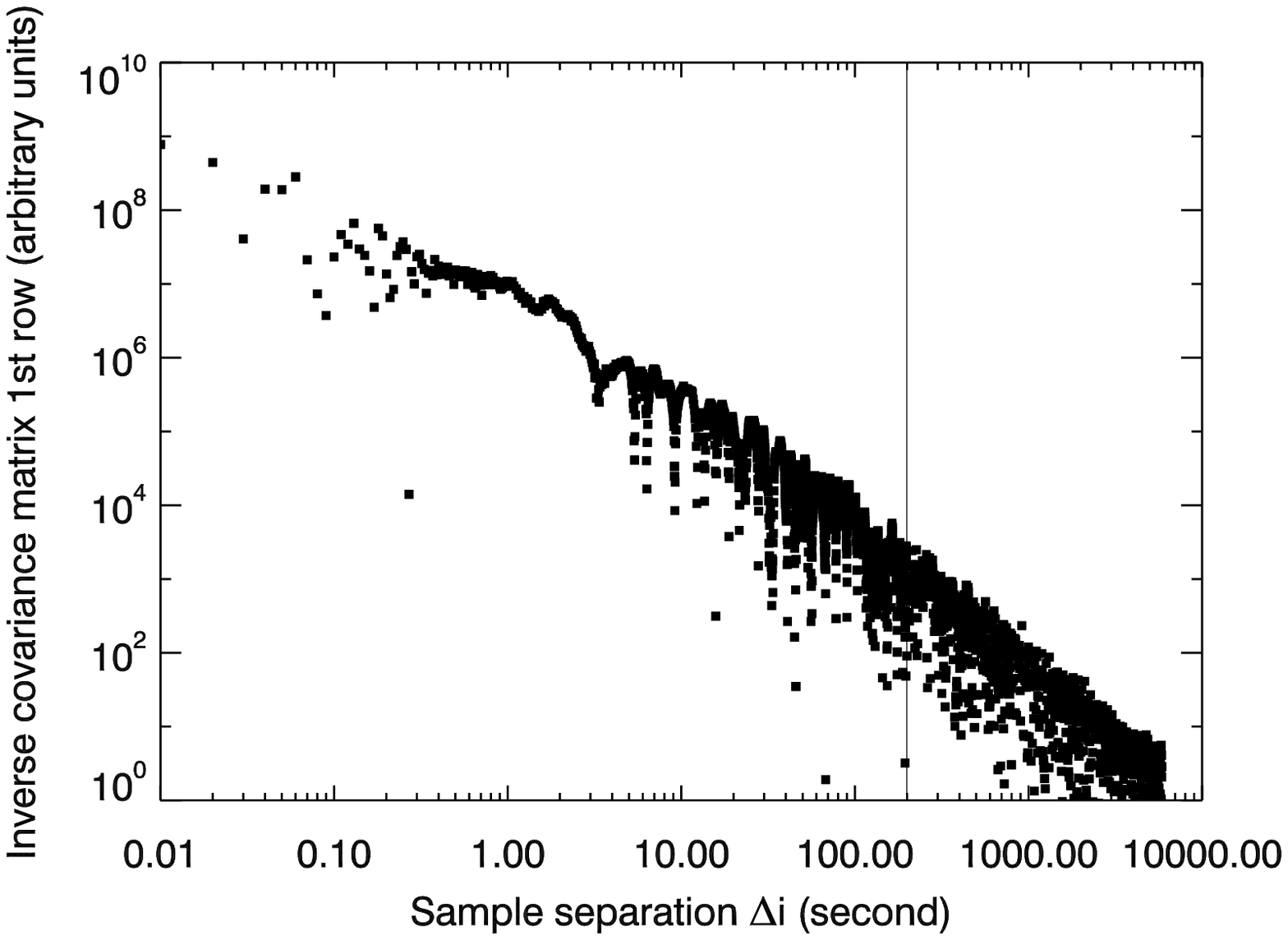}
    \includegraphics[width=\columnwidth]{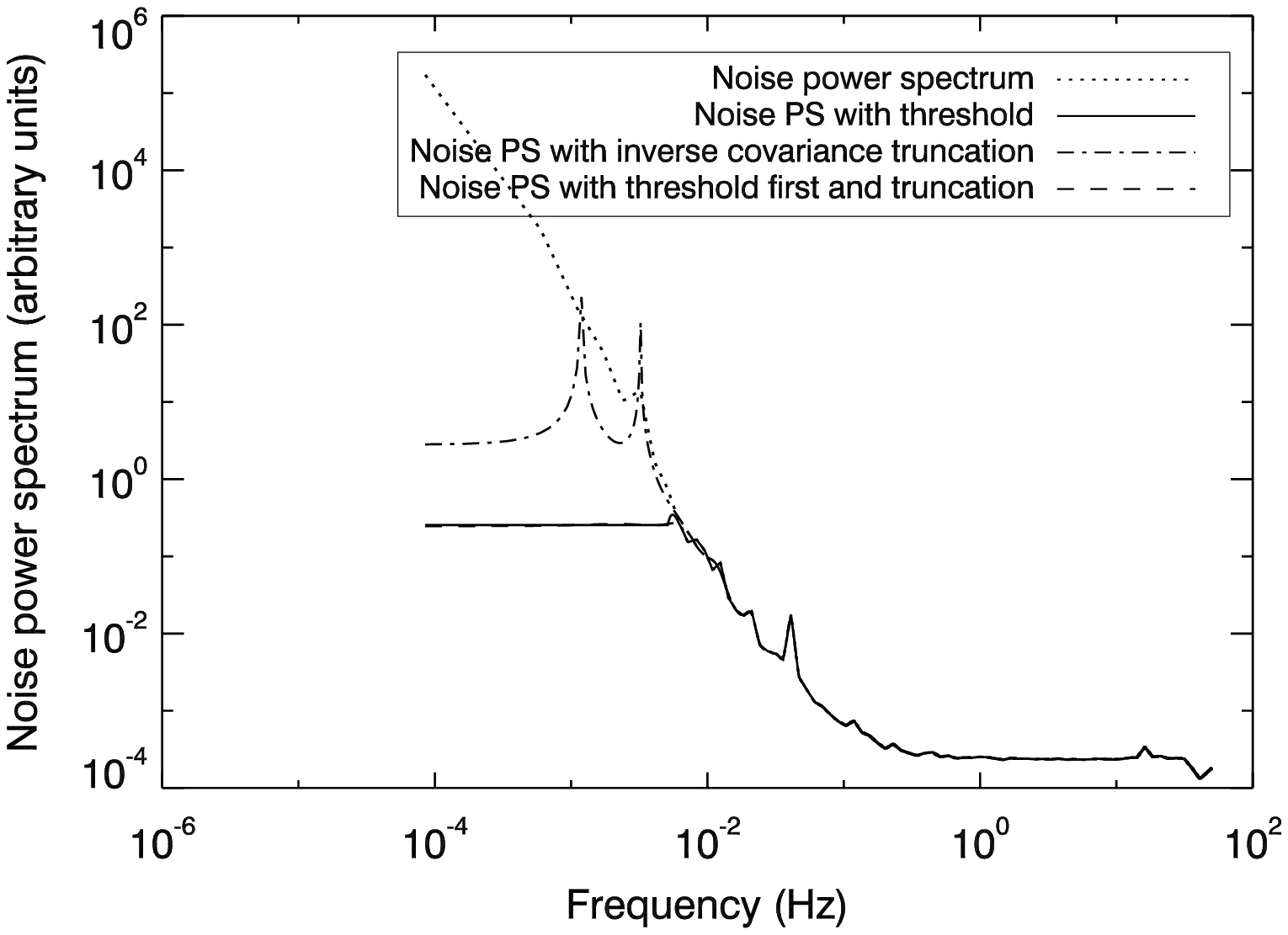}
  \end{center}
  \caption{Left panel: Absolute value of the first row of the inverse
    covariance matrix given by $\bar{C}(\Delta t)$ (see
    Section~\ref{sub:NN}) for a typical BLAST observation.  The
    vertical line indicates the value of $\lambda'_{\rm c}$, such that
    for $\Delta t > \lambda'_{\rm c}$, $\bar{C}(\Delta t)$ is set to
    zero for the computation of $N^{-1}$.  Right panel: Power spectrum
    of the noise in a typical BLAST observation (dotted curve), and
    after thresholding at frequencies smaller than $5\times
    10^{-3}\,$Hz, i.e.,~1/$\lambda'_{\rm c}$ (solid curve).  The
    dot-dashed curve is obtained by inverting Equation~\ref{eq:PktobC}
    after forcing the values of $\bar{C}(\Delta t)$ to zero for
    $\Delta t > \lambda'_{\rm c}$, with $\bar{C}$ obtained from the
    initial power spectrum of the noise (dotted curve).  To find the
    power spectrum which corresponds to the dashed curve the same
    procedure is applied, but starting from the thresholded power
    spectrum (solid curve).  All the curves are very similar for
    frequencies larger than 1/$\lambda'_{\rm c}$, but the dot-dashed
    curve begins to diverge at smaller frequencies, while the dashed
    and solid curves lie very close to each other for all frequencies.
    This shows the tight relation between the power spectrum and the
    inverse covariance matrix.  The noticeable peak in the power
    spectrum is located at the scanning frequency of about $4\times
    10^{-2}\,$Hz.}
  \label{fig:invc}
\end{figure*}

Unfortunately, if the noise is described by a power spectrum
of the form $(1/f)^\beta$, the correlation length of the noise is
basically of the order of the length of the whole data-set.  However,
the amplitude of the correlation is decreasing for very long
timescales and becomes negligible beyond a certain scale.  The
function $\bar{C}(\Delta t)$ can then be artificially set to zero
for $\Delta t > \lambda'_{\rm c}$, with $\lambda'_{\rm c}$ chosen
empirically so that the correlation of the noise is low enough for
scales longer than $\lambda'_{\rm c}$, and also that there is very
little constraint on the signal at those scales.  For the specific
case of BLAST observations, we find that $\lambda'_{\rm c}\simeq
200\,$s is a good compromise, as illustrated in Figure~\ref{fig:invc}.

The noise strongly dominates the signal in BLAST observations for
scales longer than $\lambda'_{\rm c}$ (corresponding to frequencies
smaller than $5\times 10^{-3}\,$Hz), since: (1) this frequency
is well below the knee frequency of the noise
power spectrum; and (2) there is very little signal at frequencies
smaller than $5\times 10^{-3}\,$Hz, which is more than ten times
smaller than the scanning frequency.  Therefore we have used
$\lambda'_{\rm c} = 200\,$s for the computation of the noise
covariance matrix.  The impact of fixing $\bar{C}(\Delta t) = 0$ for
$\Delta t > \lambda'_{\rm c}$ in the initial power spectrum is shown
in Figure~\ref{fig:invc}.  We can see a tight relation between the
power spectrum and the inverse covariance matrix, since getting
$\bar{C}(\Delta t) = 0$ for $\Delta t > \lambda'_{\rm c}$ can be
obtained only by modifying the power spectrum at frequencies smaller that
1/$\lambda'_{\rm c}$.

\subsubsection{Computation of $A^t N^{-1}d$}
\label{subsub:ANd}

In the computation of $x = A^t N^{-1}d$, which is necessary for both of our
algorithms, the multiplication $N^{-1}d$ is performed in Fourier space
where the noise covariance matrix is diagonal.  We obtain this vector
by dividing the Fourier transform of the data by the power spectrum of
the noise.  Another way to represent this is by considering that since
$N^{-1}$ is circulant, $N^{-1}d$ is a convolution operation. Assuming
that this model holds, the resulting data vector $\tilde{d}
= N^{-1}d$ contains whitened noise. The remaining operation $A^t \tilde{d}$
just performs the addition of the filtered data sample onto the pixels of
$s$ (i.e.~the map), and hence is very fast.

In the case of BLAST observations, since we are not attempting to recover
useful information from timescales larger than $200\,$s, we
perform a high-pass pre-filtering of  the input data $d$ at
$5\times 10^{-3}\,$Hz.

\subsubsection{Matrix inversion algorithm}

In the matrix inversion algorithm, $N_{pp'}^{-1}$ is directly computed,
as described in Section~\ref{subsub:InvNpp}. The next step is to solve 
the linear system ${\hat s} = [N_{pp'}^{-1}]^{-1}x$.
For small maps, in which $N_{pp'}^{-1}$
can be stored in memory, we perform a Cholesky decomposition.  For
larger maps, we write the matrix to disk and perform an iterative
inversion of the system using a conjugate gradient method with
pre-conditioner.

This algorithm allows one to easily perform multiple Monte-Carlo simulations,
since the pixel-pixel covariance matrix, being independent on the data
realization, can be computed once (this is assuming that the noise
power spectrum is known and does not have to be estimated from the data
themselves),
and used for all the realizations of simulated data. Another advantage
of this approach is that it allows for the exact computation of the
errors in the map, which are given by the covariance matrix.

However, the matrix inversion algorithm is generally slower than the
iterative algorithm which we will discuss next,
and can be used only for relatively small maps; we found that we were
limited to around 200{,}000 pixels if the matrix
is written to disk, or less than 20{,}000 pixels if the matrix is
stored in memory.

\subsubsection{Iterative algorithm}
\label{subsub:IterAlgo}

We now present an iterative algorithm based on conjugate gradient with
pre-conditioner to obtain the Maximum Likelihood solution for the map
(a similar algorithm has been used in \citet{ashdown}).  Let us
rewrite Equation~\ref{eq:solmap}, which relates the best estimate of
the map with the data, after multiplying both sides by the pixel-pixel
covariance matrix:
\begin{equation}
  A^tN^{-1}A~\hat{s} = A^tN^{-1} d.
\end{equation}
If we define $\hat{s}_k$ as an estimate of the map at iteration $k$,
the conjugate gradient method allows to solve the linear system by
minimizing iteratively the following criterion:
\begin{equation}
  \Psi = r^t N_{pp'}^{-1}r,
\end{equation}
where
\begin{equation}
  r \equiv (A^tN^{-1}A~\hat{s}_k - A^tN^{-1} d).
\end{equation}
This criterion is indeed minimum and equal to zero if $\hat{s}_k$ is
the Maximum Likelihood solution.

One can interpret $\Psi$ as the weighted variance of the difference
between two map vectors.  The first of these vectors, $A^tN^{-1}A~
\hat{s}_k$, is the inverse pixel-pixel covariance matrix times the
current estimate of the map, while the second vector, $A^tN^{-1} d$,
is a map constructed by simply co-adding pre-whitened data. We decide
that convergence is reached when the quantity $r^tr$ (which is much
easier to compute than $\Psi$, and also converges to zero) gets
smaller than a predefined value.  In practice the number of iterations
required for convergence is of the order of 100.

The conjugate gradient method is not described here, since it
is a fairly standard numerical tool, and the interested reader can find
many descriptions in the literature.\footnote{\scriptsize{e.g.~the 1994 article by
J.R. Shewchuk: {\tt http://www.cs.cmu.edu/ \~{}quake-papers/painless-conjugate-gradient.pdf}}}
Instead of describing the details, we focus here on aspects which are
specific to our map-making process.  In particular,
let us describe the computation of $A^tN^{-1}A~\hat{s}_k$, which is the
time-consuming part of the optimization and has to be performed at
each iteration (the computation of $A^tN^{-1} d$, also time consuming,
but needs to be done only once, since none of the parameters are changing
through the iterations); the other operations for updating the map at each
iteration are significantly faster.

One advantage of this iterative algorithm is that the computation of the full
pixel-pixel covariance matrix is not required, and the operation
$A^tN^{-1}A~\hat{s}_k$ can be done step by step.  Indeed, we start by
computing $\hat{d} = A \hat{s}_k$, which is an estimate of a ``signal''
timestream.  This operation is equivalent to scanning over the current
estimate of the map using the pointing solution.  The subsequent
operation $A^tN^{-1} \hat{d}$ (which should now be familiar), is carried out
in Fourier space,
as described in Section~\ref{subsub:ANd} (without applying any extra filtering),
and in Section~\ref{sub:detdetcorr} for the case of correlated noise between
detectors.

This iterative approach is in general much faster than the brute-force
inversion approach, because the most time-consuming operations are performed in
Fourier space.  It also requires less memory, since $N_{pp'}^{-1}$ is not
explicitly computed.  Of course, if there are found to be (or known to be)
non-trivial correlations in $N_{pp'}^{-1}$, then it may have to be
calculated explicitly, hence requiring the brute-force approach.  However,
provided the pixel-pixel correlations only involve relatively few pixels,
it should be possible to calculate a restricted part of (or perhaps
an approximation for) $N_{pp'}^{-1}$ in a modified iterative approach.
A related concept is discussed in Section~\ref{sub:errorestim}.

\subsection{Multi-scan, multi-detector case}

In the previous sub-section, we presented the general method for the
simple case where only a single continuous observation is considered.
We now describe how we combine observations from different detectors
at the same wavelength, as well as different data segments obtained
over different ``visits'' during the flight, where by ``visit'' we mean a
period in the data which starts after a sufficiently long gap, or
after the observation of a different region of the sky.

For convenience, the data vector $d$ in our model
(Equation~\ref{eq:linmodel}) now contains all the individual data
segments from different detectors, and also within a single channel,
concatenated end to end. The noise vector $n$ in
Equation~\ref{eq:linmodel} is defined in a similar manner. The matrix
$A$ in Equation~\ref{eq:linmodel} is then the result of stacking
individual pointing matrices.  The Maximum Likelihood solution is also
written as in Equation~\ref{eq:solmap}, with $N$ becoming the full
covariance matrix of the noise, including all the channels and data
chunks.  To start with, let us assume that there is no correlation of
the noise between data segments.  This is a very good assumption if we
consider data segments obtained over different visits, but is
certainly not a good approximation for segments obtained
simultaneously with different channels, since we found a very strong
common-mode noise between detectors.  We will consider the simple
no-correlation case first, and then in the next sub-section
(Section~\ref{sub:detdetcorr}) we will generalize the map-making
method to account for noise correlations from different detectors.

In the absence of correlations between data segments, the time-time noise
covariance matrix $N$ is block-diagonal and each block can be inverted
separately.  Defining $N_\ell$ as the sub-covariance matrix for the data in
segment number $\ell$, and $A_\ell$ as the sub-pointing matrix going from the
map to the data segment $\ell$, the inverse pixel-pixel covariance matrix
can be written as
\begin{equation}
  N_{pp'}^{-1} = \sum_\ell A_\ell^t N_\ell^{-1} A_\ell.
\label{eq:combscan}
\end{equation}
The computation time for obtaining this matrix is proportional to the
number of data segments.  In this simple case the computation of $x =
A^tN^{-1}d$ can be written
\begin{equation}
  x = \sum_\ell A_\ell^t N_\ell^{-1} d_\ell,
\end{equation}
where the computation of each term is fast and can be performed partly
in Fourier space.

\subsection{Detector-detector correlated noise}
\label{sub:detdetcorr}

We now allow for the presence of correlations in the noise between
different detectors.  In the case of multiple visits, the noise
covariance matrix, $N$, still has null cross-terms for samples from
two different data visits. Therefore, if the data vector is sorted by
visit, then $N$ is block diagonal and each block contains the
correlation coefficients between all the detectors for the samples
within the time interval defined as a single visit.  Each visit can be
treated independently, since the sub-matrices can be inverted
separately, and Equation~\ref{eq:combscan} is still valid, but in this
case $\ell$ is the label for the blocks in $N$.  In the following, we
therefore focus on a single visit, and consider observations by all
the detectors; the generalization to multiple visits should be clear.

To simplify the notation, let $N$ denote the noise covariance matrix
for the visit being considered, with $d$ and $n$ being the data and
noise vectors, respectively, containing the timestream segments for
all the detectors put end to end, and $N_{ij}$ being a block of $N$ of
size $n_{\rm s} \times n_{\rm s}$, corresponding to the noise
correlations between detectors $i$ and $j$.  Let us define ${\bar{F}}$
the multi-channel Fourier transform operator such that
\begin{equation}
  {\tilde{n}} = {\bar{F}}n,
\end{equation}
with ${\tilde{n}}$ containing end-to-end Fourier transforms of
each data segment.  ${\bar{F}}$ is a block-diagonal matrix, and each
block is the Fourier transform operator $F$ for one data segment.

In Fourier space, the noise covariance matrix $R$ can be written
\begin{equation}
  R = {\bar{F}} N {\bar{F}}^\dagger.
  \label{eq:FourCov}
\end{equation}
If we consider a single block of the noise covariance matrix for
detectors $i$ and $j$, we obtain:
\begin{equation}
  R_{ij} = F N_{ij} F^\dagger.
\end{equation}
Under the assumption that the data are stationary and continuous at
the edges (see Section~\ref{sub:implem} for a discussion), $N_{ij}$ is
a circulant matrix, since each element $[N_{ij}]_{tt'}$ depends only
on the time interval $|t-t'|$.  $R_{ij}$ is then a diagonal matrix
with the diagonal given by the cross-power spectrum of the noise
between detectors $i$ and $j$:
\begin{eqnarray}
\nonumber
[R_{ij}]_{\omega\omega'}  & = & P_{ij}(\omega) ~~~~{\rm if}~(\omega = \omega')\\
  & = & 0 ~~~~~~~~~~~{\rm {otherwise}.}
  \label{eq:CovRP}
\end{eqnarray}
Here $P(\omega)$ is the noise covariance matrix of size $n_{\rm d}
\times n_{\rm d}$ for a given mode $\omega$, where $n_{\rm d}$ is the
total number of detectors.  The computation of the inverse of $R$ is
straightforward, since each Fourier mode can be treated independently.
If $P^{-1}(\omega)$ is the inverse noise covariance matrix for mode
$\omega$, the same relation as in Equation~\ref{eq:CovRP} applies
between $R^{-1}$ and all $P^{-1}$.

\begin{figure}[t!]
  \begin{center}
    \includegraphics[width=\columnwidth]{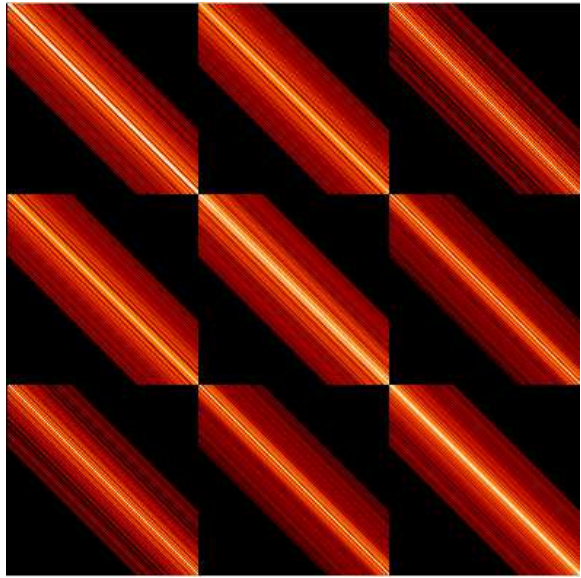}
  \end{center}
  \caption{Inverse noise covariance matrix ($N^{-1}$ in the text) for
    three detectors and for only 10 minutes of data (corresponding to
    60{,}000 samples per detector).  The matrix is computed following
    Equation~\ref{eq:InvNCorr}, and the approximations described in
    Section~\ref{sub:detdetcorr} are applied.  The cross- and
    auto-power spectra of the noise, used for the calculation of
    $N^{-1}$, are computed from the data themselves (one of the
    auto-power spectra is shown in Figure~\ref{fig:noiseSP}, the
    cross-power spectra are an unbiased measure of the common-mode
    signal shown in the same figure).  Each sub-matrix is circulant
    and corresponds to a particular pair of detectors.  One can see
    that the off-diagonal sub-matrices have amplitudes of the same
    order as that of the diagonal sub-matrices.  This is due to the
    very high level of noise correlation between detectors.}
  \label{fig:mapInvN}
\end{figure}

From Equation~\ref{eq:FourCov}, we can calculate the inverse
covariance matrix of the noise in real space:
\begin{equation}
  N^{-1} = {\bar{F}}^\dagger R^{-1} {\bar{F}}.
\end{equation}
Then, a block of $N^{-1}$ between detectors $i$ and $j$ can be written
\begin{equation}
  [N^{-1}]_{ij} = F^\dagger [R^{-1}]_{ij} F.
\end{equation}
Because $[R^{-1}]_{ij}$ is a diagonal matrix (as discussed previously)
$[N^{-1}]_{ij}$ is circulant, and is related to the inverse of the
matrix containing the cross- and auto-power spectra of the noise:
\begin{equation}
  [N^{-1}]_{ijtt'} = {\cal F}^{-1}\left\{ [P^{-1}]_{ij}\right\}(t' - t).
  \label{eq:InvNCorr}
\end{equation}
From this relation, we can see that in practice $N^{-1}$ is relatively
easy to construct, since each of its blocks (referring to each pair of
detectors) is a circulant matrix, so only a row of each sub-matrix
needs to be calculated using the Fast Fourier Transform.  Finally, in the
case the noise covariance matrix is used for multiplication in real space
(as in the brute-force algorithm), the same approximation
described in Section~\ref{sub:implem} is performed on each block of
$N^{-1}$, i.e., $[N^{-1}]_{ijtt'}=0$ for $|t - t'|>\lambda'_{\rm c}$
(or for $|t - t'|>n_{\rm s}/2$ if $\lambda'_{\rm c}>n_{\rm s}/2$).  The
global structure of the final inverse noise covariance matrix is
illustrated in Figure~\ref{fig:mapInvN}.

In our model of the data (Equation~\ref{eq:lincmod}) in which all the
correlations between detectors are described by a single common-mode,
$P^{-1}(\omega)$ can be related to the power spectra of the
common-mode and of the uncorrelated part of the noise between
detectors:
\begin{equation}
  P^{-1}(\omega) = \left(\alpha \left\langle c(\omega)^\dagger
  c(\omega)\right\rangle \alpha^t
  + \left\langle{\tilde{n}}(\omega)^\dagger{\tilde{n}}(\omega)\right\rangle
  \right)^{-1}.
  \label{eq:modSpCorr}
\end{equation}
This relation can be generalized if multiple common-mode
components are present: $\alpha$ would then be a mixing matrix and
$\left\langle c(\omega)^\dagger c(\omega)\right\rangle$ becomes the covariance
matrix of these noise components.

Having computed the inverse noise covariance matrix of the
timestreams, we can express (using Equation~\ref{eq:InvNpix})
the noise covariance matrix in the pixel domain:
\begin{equation}
  N_{pp'}^{-1} = \sum_{ij} A_i^t~[N^{-1}]_{ij}~A_j,
\end{equation}
where, as before, $i$ and $j$ label the detectors.  The computation
time for $N_{pp'}^{-1}$ is now proportional to the number of detectors
squared.  The calculation of $x=A^tN^{-1}d$ is straightforward:
\begin{equation}
  x = \sum_{ij} A_i^t~[N^{-1}]_{ij}~d_j,
\end{equation}
and this is fast, since (as already shown) the operation
$[N^{-1}]_{ij} d_j$ is a convolution, which can be performed in Fourier
space.  One can see from Equation~\ref{eq:InvNCorr} that $x$ can be expressed
directly with respect to the cross- and auto-power spectra of the
noise:
\begin{equation}
  x = \sum_{ij} A_i^t~{\cal F}^{-1}\left\{ [P^{-1}]_{ij}(\omega) .
 \bar{d}_j(w)\right\}.
\end{equation}

The formalism presented above can also be generalized easily to deal
with detectors operating at different wavelenths. The map vector $s$
could be merging different maps at different wavelengths and the noise
matrix $N$ would account for all the correlations of the noise between
detectors. The joint multi-band map-making would be suitable in
practice when some contaminations from thermal fluctuations in the
instrument or atmospheric emission are present, because they correlate
the noise at all wavelengths.

\subsection{Noise power spectrum}
\label{sub:NSP}

The Maximum Likelihood solution for the final map depends on the
noise power spectra for each data-set (through $N$ in
Equation~\ref{eq:solmap}), which are assumed to be perfectly known.
However, in practice the noise power spectra have to be inferred from
the data themselves and some uncertainties are associated with this
iterative process.

In practice a first (approximate) estimate of the noise power spectrum
can be obtained by rebinning the power spectra of each data segment,
neglecting the contribution of the astrophysical signal in the
timelines.  Indeed, for most of the fields observed with BLAST, and in
particular for blank extragalactic fields (like the ELAIS-N1 field in
BLAST's flight from Sweden), the noise is highly dominant over the sky
signal at all frequencies.  However, this is not true for measurements
of bright regions in the Galactic Plane.  Therefore, for the first
iteration's noise estimate we focus on the data taken for about 6 hours
while scanning the deepest extragalactic field (ELAIS-N1) and use this to
estimate the noise power spectra, which then become the noise input
for making the first set of maps of {\it all\/} of our fields.  This
approach is not entirely satisfactory, since the noise is not
stationary during the flight -- the noise power spectra are seen to
vary over long timescales, although they are quite constant within a
single visit of each field.  This non-stationarity appears most
obvious when there are variations of the scanning strategy between
different visits (a change of the scanning frequency induces a
variation of the location of peaks in the power spectrum), but can
also occur due to variations of detector loading or the detector bias
being changed during the flight.

Because of the observed non-stationarity of the noise, we would like
an independent estimate of the noise power spectra for each visit for
each of the observed fields.  After starting from a first estimate of
the noise power spectra based on ELAIS-N1 data as described above, we adopt
an iterative approach between maps and noise power spectra.  At each
iteration, the estimated maps are subtracted from the data, prior to
noise power spectrum estimation.  We can summarize our procedure in the
following steps:
\begin{itemize}
\item Estimate $P_0(\omega)$ from $d$ using a field known to have
  little signal;
\item Compute ${\hat s}$ from Equation~\ref{eq:solmap} (and also
  Equation~\ref{eq:InvNCorr} or \ref{eq:PktobC}, depending on the noise
  correlation being considered) using $P_0(\omega)$ as input;
\item Estimate $P(\omega)$ from $d - A{\hat s}$;
\item Re-estimate ${\hat s}$ from Equation~\ref{eq:solmap} using
  $P(\omega)$, and iterate on these last two steps until convergence
  is achieved.
\end{itemize}
We stop iterating when the noise power spectra do not vary by more
than 1 per thousand from iteration to iteration (and find that in
practice only 3 to 6 iterations are necessary to reach convergence).

We now focus on how we estimate the noise power spectra $P(\omega)$ in
steps 1 and 3.  For the simplest case, where no correlations are
assumed between detectors, we simply compute a bin-averaged power
spectrum for each data segment:
\begin{equation}
  P_\ell(q) = {1 \over n_q}\sum_{\omega_{{\rm min}(q)}}^{\omega_{{\rm max}(q)}}
 {\tilde d}_{\ell\omega}^* {\tilde d}_{\ell\omega},
\label{eq:binavP}
\end{equation}
where, for bin number $q$, $~n_q=\omega_{\rm max}(q) - \omega_{\rm
  min}(q) +1$, $\ell$ labels the data segment, and ${\tilde d}$ is the
data vector from which an estimate of the map has already been subtracted.
We have
chosen logarithmic spacing between bins and an estimate of $P(\omega)$
for each $\omega$ mode is obtained by logarithmic interpolation, which
leads to a smooth power spectrum estimate.

\begin{figure}[t!]
  \begin{center}
    \includegraphics[width=\columnwidth]{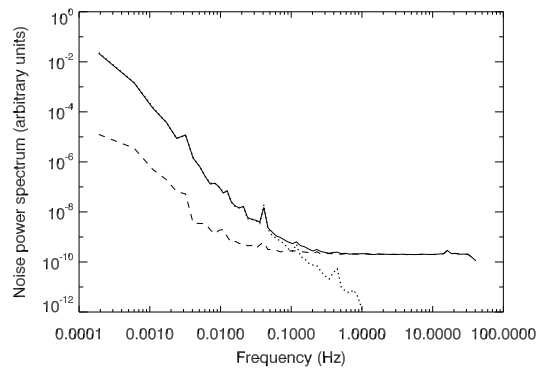}
  \end{center}
  \caption{Total noise power spectrum (solid) for one of
    the $250\,\mu$m detectors, using about three hours of data,
    corresponding to scans of the ELAIS-N1 field.  The dotted curve
    correspond to the estimated power spectrum of the common-mode
    between detectors at $250\,\mu$m, rescaled by an amplitude factor
    for the specific detector being considered ($\alpha_i$ in the
    text, where $i$ is the detector index), also estimated using the
    data themselves.  The dashed curve represents the estimated power
    spectrum of the uncorrelated part of the noise for this detector.
    The common-mode is very strong in these data and dominates over the
    uncorrelated noise at frequencies lower than about $0.1\,$Hz.
    Most of the low frequency noise excess comes from this
    common-mode.  The uncorrelated part of the noise shows a knee
    frequency of less than $0.02\,$Hz, which is 5--10 times smaller
    than for the total noise power spectrum.  The power of the
    uncorrelated noise is thus reduced by a factor of about 100 at low
    frequencies.  The excess signal at the scanning frequency (peak
    around $0.04\,$Hz in the power spectrum) is completely common
    between detectors. }
  \label{fig:noiseSP}
\end{figure}

In the more complicated case where correlations between detectors are
assumed to be important and therefore are not neglected, we must
estimate, for every iteration, each cross- and auto-power spectrum of
the data between detectors, $P_{ij}(\omega)$, which enter into the
computation of the inverse noise covariance matrix
(Equation~\ref{eq:InvNCorr}).  Each cross-power spectrum could be
directly estimated as in Equation~\ref{eq:binavP} (using ${\tilde
  d}_{i\omega}$ and ${\tilde d}_{j\omega}$ in the formulae for
detectors $i$ and $j$), but instead we choose to reduce the number of
parameters to estimate at each step, by assuming that the data are
described by a common mode between detectors plus {\it independent\/} noise
(Equation~\ref{eq:lincmod}).  In the framework of this model, the
expected cross- and auto-power spectra depend directly on the
following parameters: $\alpha$, the amplitude of the common mode in
each channel; $\left\langle c(\omega)^*\cdot c(\omega)\right\rangle$,
the power spectrum of the common-mode part; and
$\left\langle{\tilde{n}}(\omega)^*\cdot {\tilde{n}}(\omega)\right\rangle$,
the power spectrum of the noise component, which is independent
between detectors.  The relation between the model of $P(\omega)$ and
the parameters has been shown in Equation~\ref{eq:modSpCorr}.  These
parameters are typically not known {\it a priori}, and must be
measured using the data themselves.  We use a blind `component separation'
method developed
for an entirely different problem in \cite{delabrouille03}.  This allows us
to obtain a single estimate of all the parameters described
previously, by simultaneously using all the observed timestreams of a
given field for all the detectors in a specified channel (i.e.~at a
single frequency).  The method is known to be the Maximum Likelihood
solution for a Gaussian and stationary model of both the noise and the
common-mode.  The cross- and auto-power spectra $P(\omega)$ are then
computed following Equation~\ref{eq:modSpCorr}, using these same
estimated parameters.  Figure~\ref{fig:noiseSP} shows the estimated
noise power spectra in a sample of BLAST data for one representative
detector using three hours of timestreams during scans of the ELAIS-N1
field (which is known to be essentially devoid of signal).
The auto-power spectrum is shown, as well as its decomposition in terms
of the common-mode power spectrum and the uncorrelated noise power
spectrum.

Because our estimate of the converged map of the sky ${\hat s}$ is not
perfect and contains contributions from residual noise, then in
subtracting a simulated signal timeline from the data to estimate the
noise power spectrum we reintroduce some noise to the data which could
potentially bias our estimate of the noise power spectrum
\citep{ferreira,hamilton}. However this effect is greatly reduced by
the large redundancy in each pixel of the final maps, as a result of
the many repeated scans and the large number of detectors at each
wavelength. The bias can be neglected to first order for BLAST, since
the noise level per pixel in the final map is much smaller than the
noise in individual detector timestreams.

\subsection{Dealing with gaps in the data}
\label{sub:gaps}

In order to derive the formalism presented so far, we have assumed
that each data segment is stationary and hence consists of a
continuous series of
data points.  However, we have seen in Section~\ref{sub:process} that
the BLAST05 data contain multiple gaps of typical size less than one
second.  The amount of data in these gaps is a few percent of the
total.  In order to reasonably restore the continuity of the data
we have filled the gaps with random noise, as described in
Section~\ref{sub:process}.
The data samples generated in the gaps are
not reprojected into the final map but are directed to ``dummy'' pixels.
In principal, the optimal approach would be to create one dummy pixel
per flagged data sample, avoiding the possibility of several simulated
samples falling on the same pixel (through the rebinning of flagged
pixels we do not want to introduce any spurious constraints for the
map-making process which could arise from adding crossings over
different time intervals).  However, the approach of using one dummy
pixel per gap sample is impractical, because the total number of dummy
pixels would be excessive for typical BLAST observations, and the
pixel-pixel covariance matrix (which has size the total number of
pixels squared) would be prohibitive to store and compute.

We have adopted a simpler approach, which does not lead to the
mathematically exact solution, but comes very close (as has been shown
in simulations).  This consists of rejecting (for the computation of
$N_{pp'}^{-1}$) all the elements of $N_{tt'}^{-1}$ associated with flagged
samples.  This is equivalent to removing from $N_{pp'}^{-1}$ the rows
and columns corresponding to the dummy pixels before the inversion of
the matrix, as opposed to after inversion, which would be the correct
treatment discussed above.  This approach is also equivalent to
assuming null off-diagonal terms for those rows and columns.  However,
such dummy pixels are obviously correlated to some degree with the real
pixels in the map, and hence this cannot be entirely correct.

Nevertheless, we have verified that this approximation has a small
impact (a few percent only) on the final map for most of the angular
scales which are sampled, although we found some differences at very
large angular scales, sizes of the order of the map size.
For now, we have not put much
effort into recovering those very large scales because they are
subject to other effects, as discussed in
Section~\ref{subsub:sigonlyIVCG}.  The minimal impact on the final map has
been verified using pure signal simulations and by comparing the
results obtained between our simple approach and the correct
map-making solution.  This approach works to a high degree of accuracy
on most scales because the gaps are small and do not introduce
important discontinuities in the timestreams.

We have used this simple approach because it gives sufficiently
accurate results over the relevant angular scales, while being simple and
fast to implement.
However, another iterative procedure could be adopted,
which would lead to the exact solution.
In this approach,
we define two maps.  The first map ``A''is made from only the uncorrupted
(i.e.~real sky) samples, while the second map ``B'' is obtained from
projecting the simulated (i.e.~for gap-filling) samples.  The difficulty
arises in deciding what to do when simulated data
from different scans or detectors fall in the same pixel -- one might want
the ``generated signal'' (and not necessarily the noise) to be identical in
both measurements, in order to satisfy the map-making hypothesis.
If this condition is not satisfied, then some artifacts may be introduced
into both maps.  Here is a solution to this problem:

\begin{enumerate}
\item generate a first set of maps ``A'' and ``B'' after filling the gaps
  in the timestreams with white noise + a linear baseline.
\item fill the gaps in the data with the best estimate of the signal in
  map ``A'' together with white noise + a baseline which is fitted in the
  gap vicinity of the data ``minus'' signal timestream.
\item recompute the maps ``A'' and ``B''.  Step 2 ensures that the signal is
  the same for each generated sample falling in the same pixel of map ``B''.
\end{enumerate}

This approach can also be coupled with the procedure for estimating the
noise power spectra described in Section~\ref{sub:NSP}.  Preliminary
results indicate that this approach works in practice.  Detailed
studies will be presented in a future publication.

\subsection{Pixel constraints}
\label{sub:pixconstr}

For some specific observed fields we may have strong priors about the
sky emission at a given location.  For instance, we know that over
some regions the astronomical signal should vary very smoothly or should be
very weak with respect to the noise, at least outside some localized
region. This is the case in particular when we map bright
extragalactic sources in order to calibrate the detectors and estimate
the beams; in these cases, regions beyond some predefined distance
from the beam center can be assumed to have null flux (or a constant
relative flux in the map, since we do not have access to the DC level
in maps).  If we really have strong prior knowledge that we are
dealing with a bright localized region, then we can take a further
drastic step -- we can constrain the map to have the same value in some
domains of the sky by defining a single pixel containing all the data
samples falling in that region.

In practice, we define a small box centered at the source location and
constrain the part of the map outside this box to have a constant
value.  This is a very efficient way to remove stripes from the map,
since the extremities of all the paths across the map are re-adjusted.
We have used this technique to make maps of the isolated calibrators
observed by BLAST (Truch et al.~2007).

\subsection{Error estimation}
\label{sub:errorestim}

The variance of the noise in each pixel of the final map and its
correlations are directly given by the pixel-pixel covariance matrix
$N_{pp'} = (A^t N^{-1} A)^{-1}$.  This is true given the following assumptions:
that our model of the data holds, in particular that the noise is a purely
Gaussian random process, which may not be the case in practice at low
frequencies; and that our estimate of the sample-sample
noise covariance matrix $N_{tt'}$ is accurate enough that the errors do not
propagate significantly into the final map.  As already mentioned, we
never explicitly compute the covariance matrix, but rather its
inverse.  The direct inversion would take a prohibitive
computation time for most applications.  However, to first order, we can
obtain an
estimate of the errors by inverting the diagonal part of $N_{pp'}^{-1}$
only, neglecting the off-diagonal terms.  This is equivalent to assuming
that the noise in the final map is white.  We have checked with the
help of simulations that this very simple approximation is accurate to
better than 10 percent for all our BLAST05 fields, even for those with very
poor cross-linking.

Provided that the size of the map is reasonably small, so that we are
able to explicitly calculate $N_{pp'}$, we can obtain an accurate estimate
of the errors for a limited (and small) number of pixels in the map.  The
variance for pixel $p$, and the covariance with respect to the other
pixels of the map, can be computed by solving the linear system:
\begin{equation}
  \left\langle n_p^\dagger \cdot n_{p'}\right\rangle = N_{p'p} u_p,
\label{eq:var}
\end{equation}
where $u_p$ is a unitary vector with a single 1 for pixel $p$. If the
Cholesky decomposition of the matrix has already been performed for
the map-making procedure, the computation of
Equation~\ref{eq:var} is relatively fast and hence can be carried out for a
grid of non-adjacent pixels, for example.  This can be used to check
the validity of the error prediction approximation described in the
previous paragraph.

\subsection{Computational requirements}

For the brute-force inversion algorithm (in which the full inverse
pixel-pixel covariance matrix $N_{pp'}^{-1}$ is computed), five
minutes of computation with a single $3\,$GHz processor are needed to
process 2 hours of data from a single detector at a rate of $100\,$Hz.
The computational time is proportional to the number of samples if
this is longer than the assumed correlation length of the noise in the
data (which has been evaluated to be $\lambda_{\rm c} = 200\,$s in
BLAST05 timestreams).  If noise correlations between detectors are
also to be accounted for, the computational time is proportional to
the square of the number of detectors.

Most of the computing time is spent on calculating $N_{pp'}^{-1}$.
Inversion of the linear system to estimate the map $s$ is relatively
fast (a few minutes to a few hours for maps of several square degrees
in size).

For the iterative algorithm, about two minutes of computational time
is required for two hours of data (under the same conditions described
above).  This assumes that 100 iterations are necessary to reach
convergence (which is a realistic number for most applications), and
the algorithm scales with $n_{\rm s}\log n_{\rm s}$. The situation is
much better than for the brut-force inversion algorithm if
correlations between detectors are included. In that case, the
algorithm scales with the square of the number of detectors if this
exceeds about 40.  If there are fewer than about 40 detectors then the
algorithm scales linearly with the number of detectors.  As an
example, if there are 100 detectors, then including noise correlations
between the detectors increases the computational time by a factor of
four with respect to the ``no-correlation'' case.  The full processing
of 10 hours of BLAST05 data at all wavelengths, including
detector-detector correlations, can be done with a single processor in
a few days.

In terms of memory, the brute-force inversion algorithm requires
storage of the full $N_{pp'}^{-1}$ matrix.  However, for the iterative
algorithm, only vectors of the size of the maps need to be kept in
memory, which is much less demanding.

\section{Simulations for testing SANEPIC}
\label{sec:appli}

We now focus on the application of SANEPIC to data. Our aim is to
develop tests to validate our method using simulated BLAST
observations. We derive conclusions about how well low frequency noise
in the maps can be reduced, depending on observational parameters such as
scanning strategy, and we compare the results obtained with those from
simpler methods based on filtering the data, e.g., common-mode
subtraction.
In this section, we describe the simulations performed to test the
SANEPIC method.

We have generated several different sets of simulations of BLAST timestreams.
Each set of simulated timestreams,
representing one particular observed field, is generated for all the
BLAST detectors used for the analysis of real data (132 at
250$\,\mu$m, 78 at 350$\,\mu$m and 39 at 500$\,\mu$m) and is the sum
of simulated astrophysical signal, independent noise, and common-mode
noise between all detectors. The noise is generated randomly with
Gaussian statistics, given fixed power spectra derived from real BLAST05
data.  Figure~\ref{fig:noiseSP} shows an example of the power spectrum
of the noise in the data used as input to the simulations for one
of the fields. The part of the noise which is independent between
detectors is generated for every detector timestream and has a power
spectrum well described on average by a relatively flat plateau for
frequencies larger than about $0.05\,$Hz, and by a part proportional
to $(1/f)^{2.5}$ for smaller frequencies (these characteristics vary
slightly from detector to detector). Our knowledge of the real
bolometer noise power spectrum at low frequency is limited by the very
dominant common mode. In simulations, the common-mode noise is
generated once for all detectors and has a power spectrum very well
fit by a power law with an index equal to approximately 2.5, together
with some broad peaks, the largest being at the scanning frequency
(the amplitude of the peak depends on the scanning strategy and the
observed field).  The common-mode power spectrum has an amplitude such
that it reaches the level of the independent noise at about $0.3\,$Hz
(see Figure~\ref{fig:noiseSP}).  The generated common-mode timestream
is multiplied by an amplitude factor which varies from detector to
detector by $\sim 10\%$ and is added to the simulated detector
timestreams. The amplitude factors used for the simulations have been
estimated from the data themselves.

In order to represent the astrophysical signal, we have simulated
simple maps of diffuse emission with a power spectrum proportional to
$k^{-3}$, as for typical Galactic cirrus emission (e.g.~Miville-Desch{\^e}nes
et al.~2007). Maps are generated
following Gaussian statistics with a resolution of 1$\arcsec$, much
higher than the typical pixel sizes in the final maps, in order to
reduce artifacts due to re-pixelization. The amplitude of the
fluctuations of the simulated map is chosen to match the expected
level of signal in each observed field. The simulated maps are scanned
using BLAST05 pointing, and pure signal timestreams are generated for
each detector. Signal and noise timestreams are added at the end of
the procedure (but see Section~\ref{sub:results} for an explanation of
why this operation is not always carried out).

We have generated two sets of simulations which correspond to two
different fields observed by BLAST. We selected two fields that were
observed with very different scanning strategies, since the
performance of the map-making procedure is very dependent on scanning
strategy; this allows us to test SANEPIC in two very different
configurations. In the first case the scanning was performed mainly in
a single direction over a short time interval, while in the second case
the field was observed several times during the flight at different
scanning angles, to achieve significant cross-linking in the map.

The first data-set uses observations of the \casa supernova remnant
emission which comprises about 20 minutes of data.  BLAST observations
of this field and derived conclusions will be described in detail in
Hargrave et al.~(in preparation).
The rectangular region mapped has a size of the order of $0.5\,{\rm deg}^2$
and was scanned two times back and forth over a
short time interval.  We have generated simulations corresponding to
all the detectors at 250$\,\mu$m (we used a total of 132 detectors).

The second data-set reproduces the observations of the intermediate
velocity cloud IVC G86.5+59.6 (hereafter `G86').  Simulations include
four different visits of the field performed during the flight at very
different time intervals (ranging from a few hours to more than a
day). Each continuous observation segment has a size which varies from
one to two hours. Two scanning directions are dominant, which form an
angle close to 50$^\circ$.  The region covered has a size of about
$2\,{\rm deg}^2$ on the sky.  Simulations for this field are performed
specifically for all the 500$\,\mu$m detectors (41 detectors used).
Similar Monte Carlo simulations at 250$\,\mu$m would have taken a factor
of 10 longer, while we believe that the conclusion would remain unchanged.

A total of 20 sets of simulations of the observations for each field
have been performed. For each set we vary the realization of the noise
and of the signal input map.  About four hours of computing time are needed
to create one realization of a full set of simulations of \ivcg with a
single processor, compared to a few minutes for the \casa simulations.
This is using the pre-computed full pixel-pixel covariance matrix,
which was also used to analyze the real data.

\section{Results from simulations}
\label{sub:results}

We now present the results obtained with SANEPIC applied to the two
sets of simulated data. In each case, we compare the final map with
other maps obtained using simpler map-making procedures.  For these tests
we have assumed that the noise power spectra are perfectly known, rather
than estimated separately from each data-set; in practice we fix the
noise power spectra to be the ones from the simulations.  We have
verified that relaxing this constraint has almost negligible change
on the final maps.

For these simulated data-sets, we have applied some pre-processing of the
timestreams before applying SANEPIC, just as we do for the real data.
We have systematically
removed a 5th order polynomial from each timestream segment and
weakly high-pass filtered the data, as described in Section~\ref{sub:process}.
Finally, for the gap-filling we flag the simulated data at the
same locations as in the real data in order to check the influence of
the flagging procedure in the final maps.

In the following we have applied SANEPIC independently to pure noise
timestreams (containing independent noise and common-mode noise, but
without simulated astrophysical signal) and to pure signal
timestreams.  This procedure allows us to easily derive conclusions
about the noise properties in the final maps, as well as about the signal,
without biasing the results, because SANEPIC is a linear method (as shown by
Equation~\ref{eq:solmap}).  This is only strictly true if the noise power
spectrum is fixed as done here, and not estimated simultaneously along
with the maps.  Then, applying SANEPIC on pure noise timestreams and
pure signal timestreams independently and adding the two final maps
is rigorously equivalent to applying SANEPIC on signal plus noise
timestreams.  This has been checked numerically, and we find that the
difference is consistent with double floating precision error.  An
important consequence of this is that the properties of the noise in
the final map are independent of the signal-to-noise ratio.

\subsection{Case without cross-linking}
\label{sub:CasACase}

\subsubsection{Noise-only timestreams}

We first study the maps resulting from the noise-only timestreams in
the configuration of \casa observations.  The chosen pixel size of the
map is 25$\arcsec$ and matches the pixel size of the maps discussed in
Hargrave et al.~(in preparation).
We compare the noise maps obtained from three different procedures:
\begin{itemize}
\item Case 1: use SANEPIC with the correct treatment of the
  correlated noise.
\item Case 2: use SANEPIC fixing the correlation of noise between
  detectors to zero and fixing the noise power spectrum for each
  detector to the power spectrum of the sum of uncorrelated noise and
  common mode. This procedure is very similar to more standard
  map-makers in the literature (e.g., Stompor et al.~2002).
\item Case 3: make a simple re-projection of the data onto a pixelized
  map by simply averaging the data falling in each pixel, after having
  filtered the timestream data with the same very weak low-pass
  filter used for SANEPIC. This procedure is sometimes called ``co-addition''.
\end{itemize}

\begin{figure*}[t!]
  \begin{center}\resizebox{16.5cm}{!}{
    \includegraphics[scale=0.35]{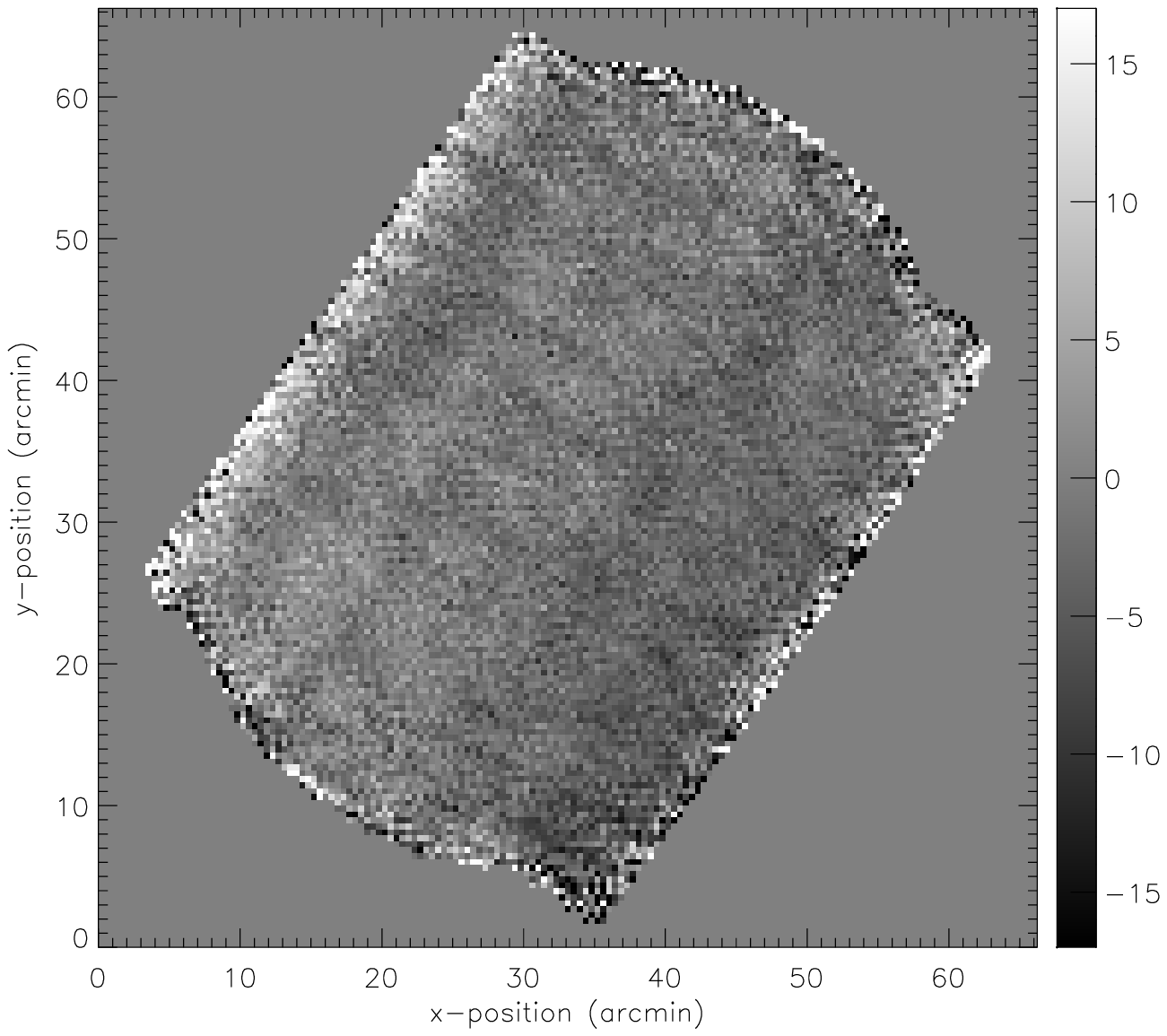}
    \includegraphics[scale=0.35]{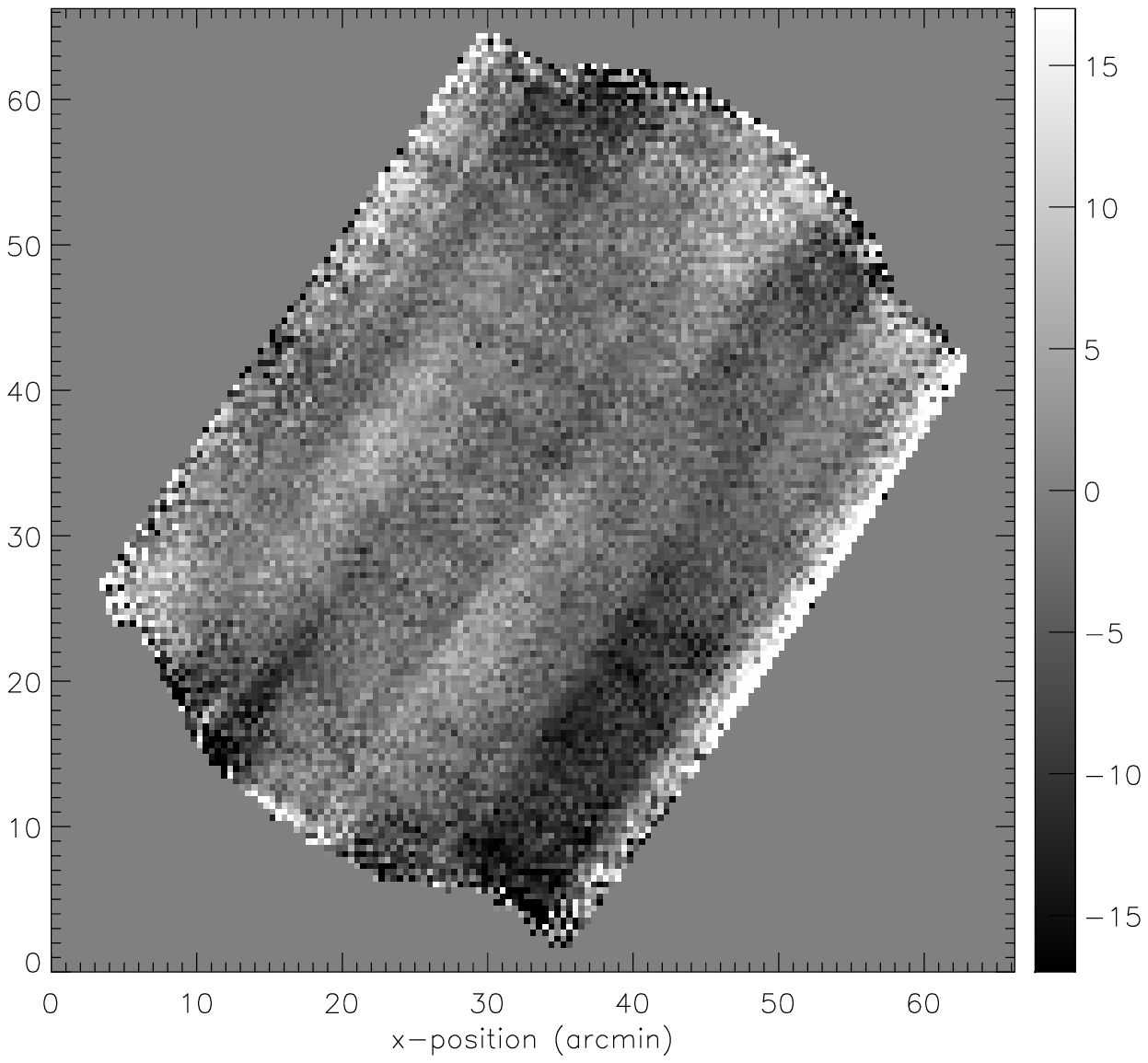}
    \includegraphics[scale=0.35]{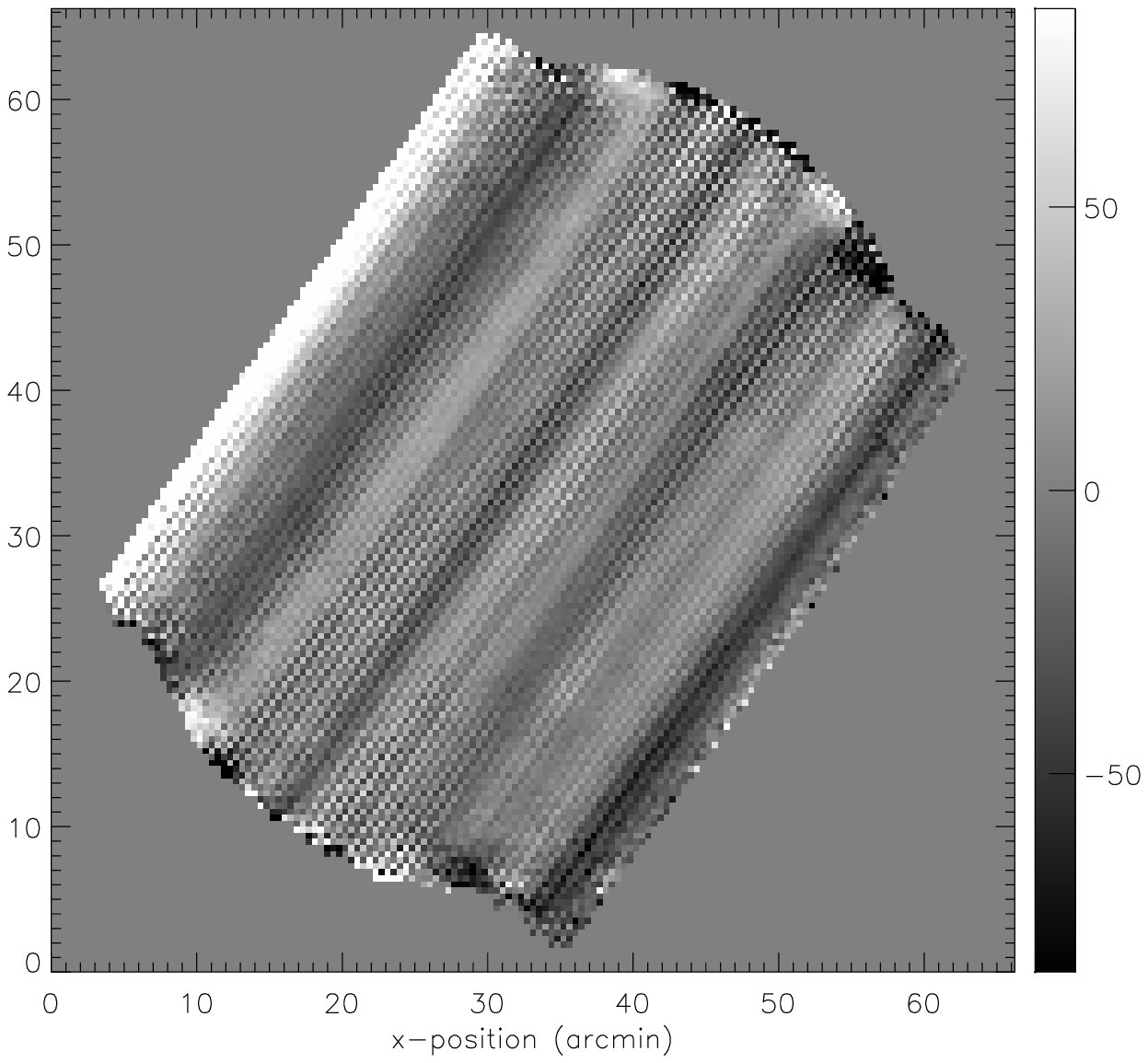}
}
  \end{center}
  \caption{Final maps computed from simulated pure noise timestreams
    in the configuration of the BLAST05 \casa observations, which
    have a dominant scan direction.  From left to right:
    maps obtained with SANEPIC including noise correlations; SANEPIC
    with no noise correlations included in the model; and simple pixel
    binning (see text for more details).  Note the extended dynamic
    range of the simple co-added map (right panel).  The maps have a
    size of about 40$\arcmin$ in the cross-scan direction and about
    $1^\circ$ along the scan.  The pixel size is 25$\arcsec$.}
  \label{fig:mapnCasA_corr}
\end{figure*}

Figure~\ref{fig:mapnCasA_corr} shows computed noise maps for one of the
realizations of the noise in each of the three cases.
As expected, the map obtained with the simple pixel binning approach
contains a very large amount of low frequency noise, with strong striping
visible along the scan direction.  Residual low frequency
noise can also be seen in the map obtained using SANEPIC {\it without\/}
accounting for the noise correlations between detectors.  We do not expect this
method to be very efficient, since it is very non-optimal in cases
(such as this example)
where a very large fraction of the noise is correlated
between detectors.  In contrast, the noise map obtained with SANEPIC is
quite satisfactory, showing reduced power at low frequency as
compared to the previous case.  Nevertheless, some very weak excess
power is seen in the cross-scan direction.  This is expected, since the
map is not cross-linked, and very poor constraints can be put on the
cross-scan directions at low spatial frequencies (two positions in the
map separated by more than the size of the array in the cross-scan
direction are observed far apart in time).

\begin{figure}[h!]
  \begin{center}
    \includegraphics[width=\columnwidth]{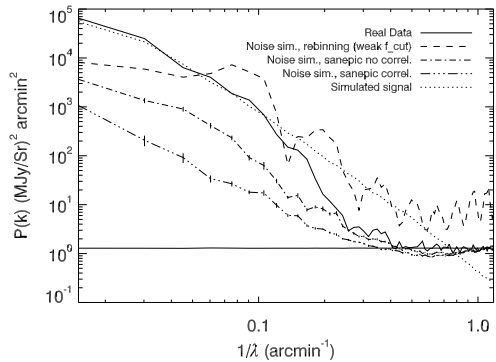}
  \end{center}
  \caption{One-dimensional power spectra of the noise (rebinned in
    frequency) in the final noise maps after map-making in the BLAST05
    \casa configuration. Power spectra are averaged over the 20
    realizations of the simulated data. The dashed curve is for the
    simple re-projection map, the dot-dashed curve for SANEPIC with
    {\it no\/} noise correlation between detectors and the
    triple-dot-dashed curve for SANEPIC {\it including\/} a treatment
    of the correlations. The straight line indicates the level of
    white noise in the map predicted by the map-making procedure (see
    Section~\ref{sub:errorestim}). Error bars are computed from the
    dispersion of measurements among the realizations.  For
    comparison, the upper dotted curve (decreasing almost like a power
    law at all scales) represents the power spectrum of the pure
    simulated signal in the final map. The solid curve
    represents the power spectrum of the final map obtained with real
    data using SANEPIC, with correlations included. This shows the
    benefit of taking into account correlations of the noise between
    detectors in the map-making procedure, reducing the noise
    structure far below that of the signal in the map. The real data
    power spectrum shows that the signal dominates at all angular
    scales larger than about 3$\arcmin$ and at smaller scales we can
    see that white noise at the expected level dominates in the
    map. The drop of power at around a 3$\arcmin$ scale is due to the
    BLAST05 beam.}
  \label{fig:noise1dSpCasA}
\end{figure}

In order to quantify the level of low frequency noise in the maps, we
compute the 1-D power spectra of the maps, averaged over the 20
realizations of the simulated data. For the computation of power
spectra, we take into account only the central part of each map, where
the level of redundancy in the observations is high (we use only the
highest signal-to-noise region in the maps).  To do so, we apply an
apodized mask to the maps going smoothly from 0 at the edges
to 1.  Figure~\ref{fig:noise1dSpCasA} shows the noise power spectra in
the three cases.
The noise level in the simple re-projection map is obviously very poor
at all scales.  Both of the other map-makers reach the white noise
level for scales smaller than 3$\arcmin$ and have excess power at
larger angular scales.  Nevertheless, the gain between full SANEPIC and
SANEPIC without correlations is very important at all scales larger than
about 2$\arcmin$ and reaches a maximum value of about 10 at around
20$\arcmin$ angular scales.  An interesting fact is that the knee
frequency of the noise power spectrum in the optimal case here corresponds
to the inverse of the physical scale of the detector array in the
cross-scan direction (which is of the order of 6$\arcmin$). Indeed,
there are no observational redundancies on scales larger than the
array in the cross-scan direction in the absence of cross-linking in
the map. Thus the very long timescale $1/f$ noise present in the
timestreams is not efficiently removed and re-projects in the final
map at large angular scales. This effect is also present along the
scan direction, but with a lower amplitude as the map is scanned back and
forth. The trend of the large angular scale power spectrum of the
noise in the map just follows the trend of the low frequency noise
power spectrum in the timestreams. We will see in Section
\ref{subsub:crosslink} that this effect is reduced when there are
multiple scanning directions in the map.

\begin{figure}[t!]
  \begin{center}
    \includegraphics[width=\columnwidth]{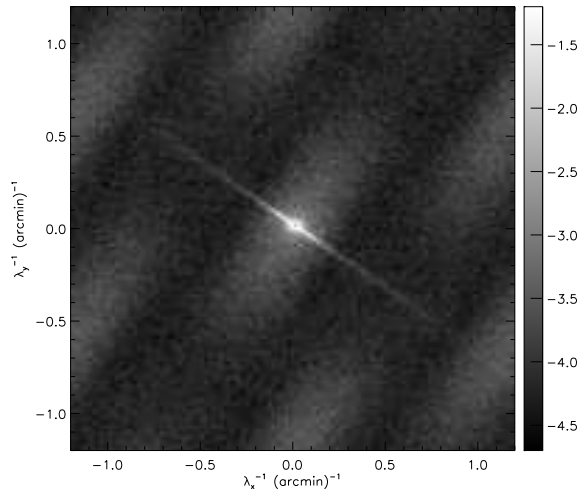}
  \end{center}
  \caption{Two-dimensional power spectrum of the noise maps in the
    BLAST05 \casa observational configuration obtained with SANEPIC (noise
    correlations included) plotted on a logarithmic contrast scale.}
  \label{fig:Sp2dnoiseCasA_corr}
\end{figure}

In order to determine the direction in which the noise power is
strongest in the map, we have also computed the 2-dimensional noise power
spectrum. The map of the 2-D power spectrum of the noise obtained with
SANEPIC (noise correlations included) is shown in
Figure~\ref{fig:Sp2dnoiseCasA_corr}.
The large bright spot around the center corresponds to a relatively
isotropic component of correlated noise (at least at large angular
scales). It contains a large fraction of the noise power at
large angular scales (seen in the 1-D power spectrum in
Figure~\ref{fig:noise1dSpCasA}).  A smaller, but significant fraction of
the correlated noise is concentrated in directions perpendicular to
the scan direction, as can be seen in the figure.  As already
discussed, the reason for this excess power is that the noise in the
cross-scan direction is poorly constrained.  This cross-scan component
of the noise is significant all the way up to the pixel scale.

\subsubsection{Signal-only timestreams}
\label{sec:CasAsignal}

We now focus on the signal-only timestream simulations. In order to
demonstrate the superior performance of SANEPIC relative to simpler
methods based on data filtering, we compare with a map-making method which
consists of the following: we first remove the whole array average
from each detector timestream and then make maps
using SANEPIC, assuming no correlations between
detectors. Removing the array average reduces the signal to almost
zero for scales larger than the array and so we expect no large scale
structures to survive in the map.  This SANEPIC ``common-mode
subtraction'' method is still a better procedure than just
reprojecting the data (after common-mode subtraction) with a well
chosen filtering to suppress noise drifts (at $f_{\rm cut} =
0.02\,$Hz, for instance, since that corresponds to the knee frequency
of the independent part of the noise). This latter method is commonly
used in the submillimeter community and is referred to as ``sky
removal'' in reduction of SCUBA data \citep{jenness}. 

\begin{figure*}[t!]
  \begin{center}\resizebox{16.5cm}{!}{
      \includegraphics[scale=0.35]{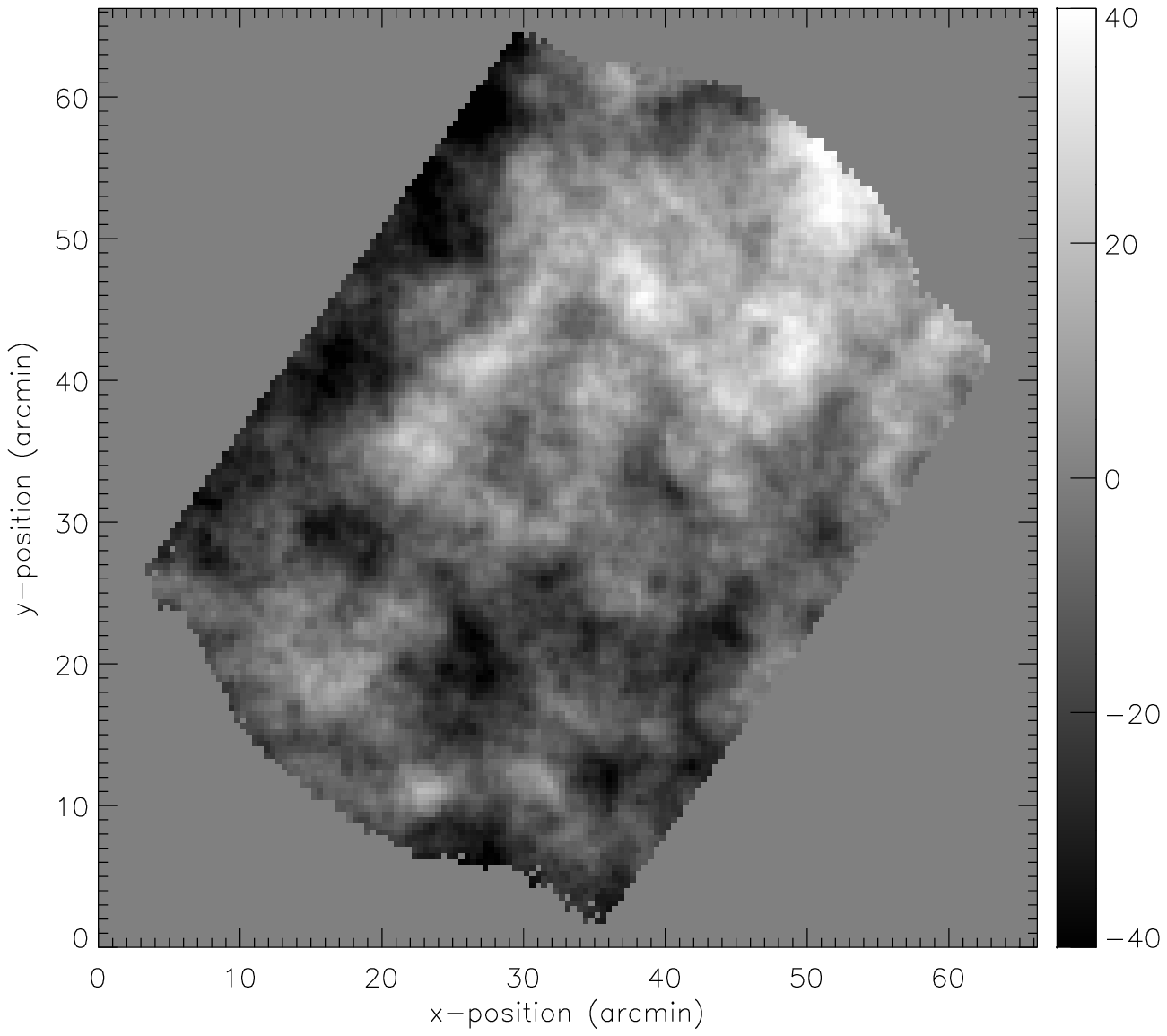}
      \includegraphics[scale=0.35]{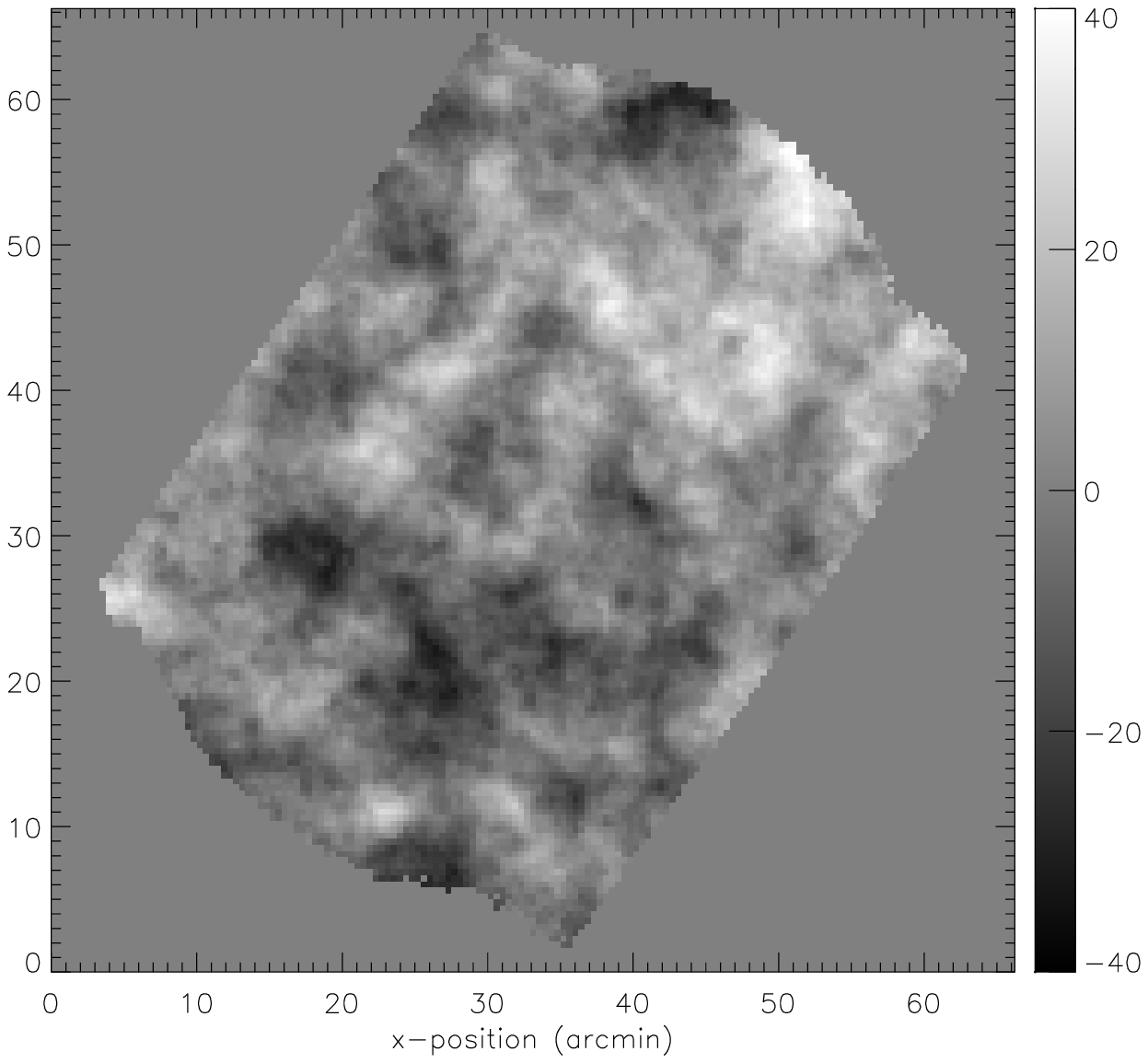}
      \includegraphics[scale=0.35]{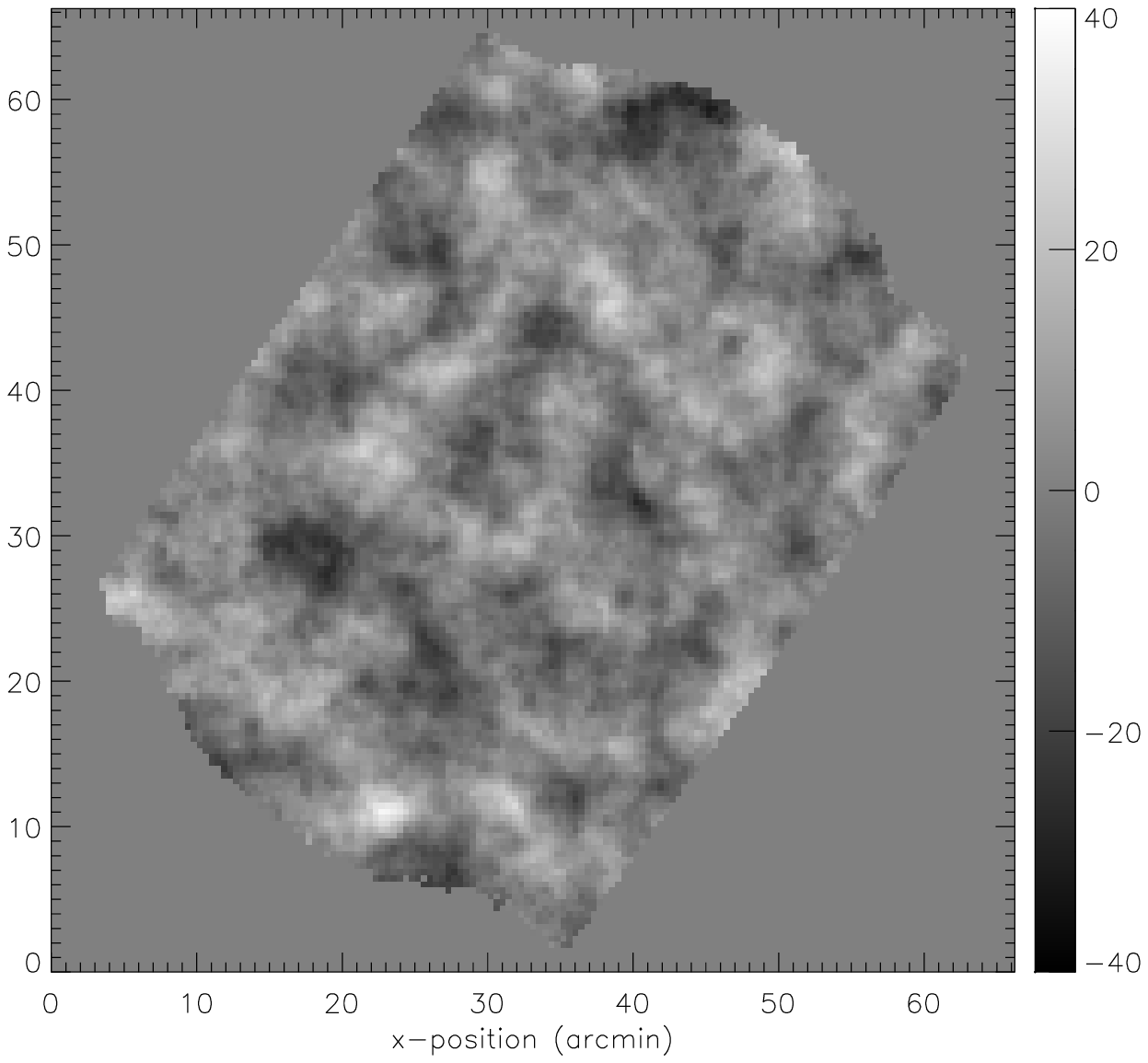}
    }
  \end{center}
  \caption{From left to right: map of simulated signal used as input
    for one of the realizations of the simulations in the
    configuration of the BLAST05 \casa field; reconstructed map using SANEPIC
    accounting for noise correlations; reconstructed map using the simple
    common-mode subtraction method (intensity units here are arbitrary).}
  \label{fig:mapsCasA}
\end{figure*}

\begin{figure}[h!]
  \begin{center}
    \includegraphics[width=\columnwidth]{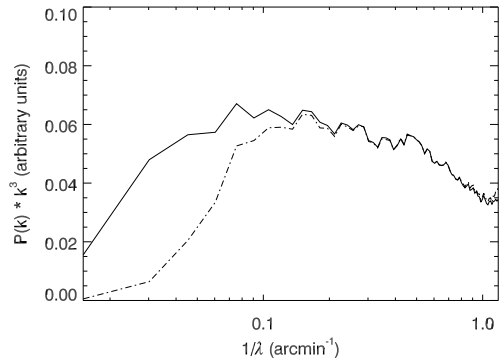}
  \end{center}
  \caption{1-D power spectra of signal-only maps for the BLAST05 \casa field
    reconstructed using SANEPIC (solid curve) and using the common-mode
    subtraction method (dot-dashed curve).  Power spectra are
    multiplied by $k^3$, displayed on a linear scale and averaged
    over 20 realizations of the simulations.  The large angular scale
    behavior shows the effectively filtering of each map-making procedure,
    while the drop off at small scales is caused by the PSF.}
  \label{fig:1DpowerCasA}
\end{figure}

Figure~\ref{fig:mapsCasA} shows the input map for one of the
signal realizations (left panel), as well as the maps obtained with
SANEPIC (correlations included, central panel) and with the common-mode
subtraction method (right panel).  Results are
expected to be worse in the second case, because of the extra
filtering and also because SANEPIC gives less weighting to modes at
lower frequency which are more contaminated by independent noise.
We can see from Figure~\ref{fig:mapsCasA}
that part of the very large scale fluctuations with sizes
of the order of the map are removed using SANEPIC, but apart from
those very large scales, the input map and the SANEPIC map look very
similar. More differences can be seen in the map obtained with the
common-mode subtraction method. This is quantified in Figure
\ref{fig:1DpowerCasA}, which compares the 1-D power spectra of the two
output maps, averaged over 20 simulations and multiplied by $k^3$.
Recall that the input spectrum varies as $k^{-3}$ and so
deviations from a flat line are the result of the map-making
reconstruction. Note that the vertical scale is linear in
Figure~\ref{fig:1DpowerCasA}.

With SANEPIC, the power of the reconstructed map decreases for scales
larger than about 30$\arcmin$.  There are three reasons for this:
the power spectrum is
computed over only a small fraction of the sky (and for an apodized map),
so that structures of the order of the map size are never fully recovered;
there is weak
filtering of the timestreams at $f_{\rm cut}=5\times 10^{-3}\,$Hz and
through the 5th order polynomial subtraction; modes in the maps that
are very weakly constrained in the map-making procedure tend not to be
reconstructed through the matrix inversion procedure, since the matrix is very
ill-conditioned and numerical problems occur. The last two effects are
the dominant ones. As a result, modes which are preferentially
filtered are those which lie perpendicular to the scan direction.

At angular scales smaller than about 2$\arcmin$, the power slightly
decreases due to the smoothing effect of the pixelization. The
transfer function at those scales is well described by a sinc
function.

Turning now to the common-mode subtraction method, the power in the map is
significantly reduced for scales larger than about 10$\arcmin$, and
drops rapidly to zero.  This is because the common-mode subtraction
removes power on all scales larger than the array.  On smaller
scales, the filtering effect is relatively weak, and is reduced when
the number of detectors increases.

For these particular simulations the common-mode subtraction method (using
SANEPIC, but with no correlations) does not in fact perform {\it very\/}
poorly compared to the SANEPIC optimal approach.
This is because the observed field is
small, with a size just a few times bigger than the array, and
structure at scales smaller than the array size are not strongly
affected.  This particular map is also not cross-linked.  However, the
situation is different for large cross-linked maps like the Vulpecula
field, as discussed in Section~\ref{sub:Vulpecula}.

\subsection{Case with cross-linking}
\label{subsub:crosslink}

\subsubsection{Noise-only timestreams}

We now focus on the set of simulations of the \ivcg field
at 500$\,\mu$m.  As in the previous example, we first examine the
maps resulting from noise-only simulated timestreams using three
methods: optimal SANEPIC (with noise correlations taken into account); SANEPIC
without considering noise correlations between detectors; and the simple
co-add method.  The chosen pixel size for the reconstructed maps is
1$\arcmin$, which allows for inversion of the covariance matrix with a
single processor and hence rapid Monte Carlo simulations.  The
conclusions drawn would remain unchanged if the pixel size was
reduced.

\begin{figure*}[t!]
  \begin{center}\resizebox{16.5cm}{!}{
      \includegraphics[scale=0.35]{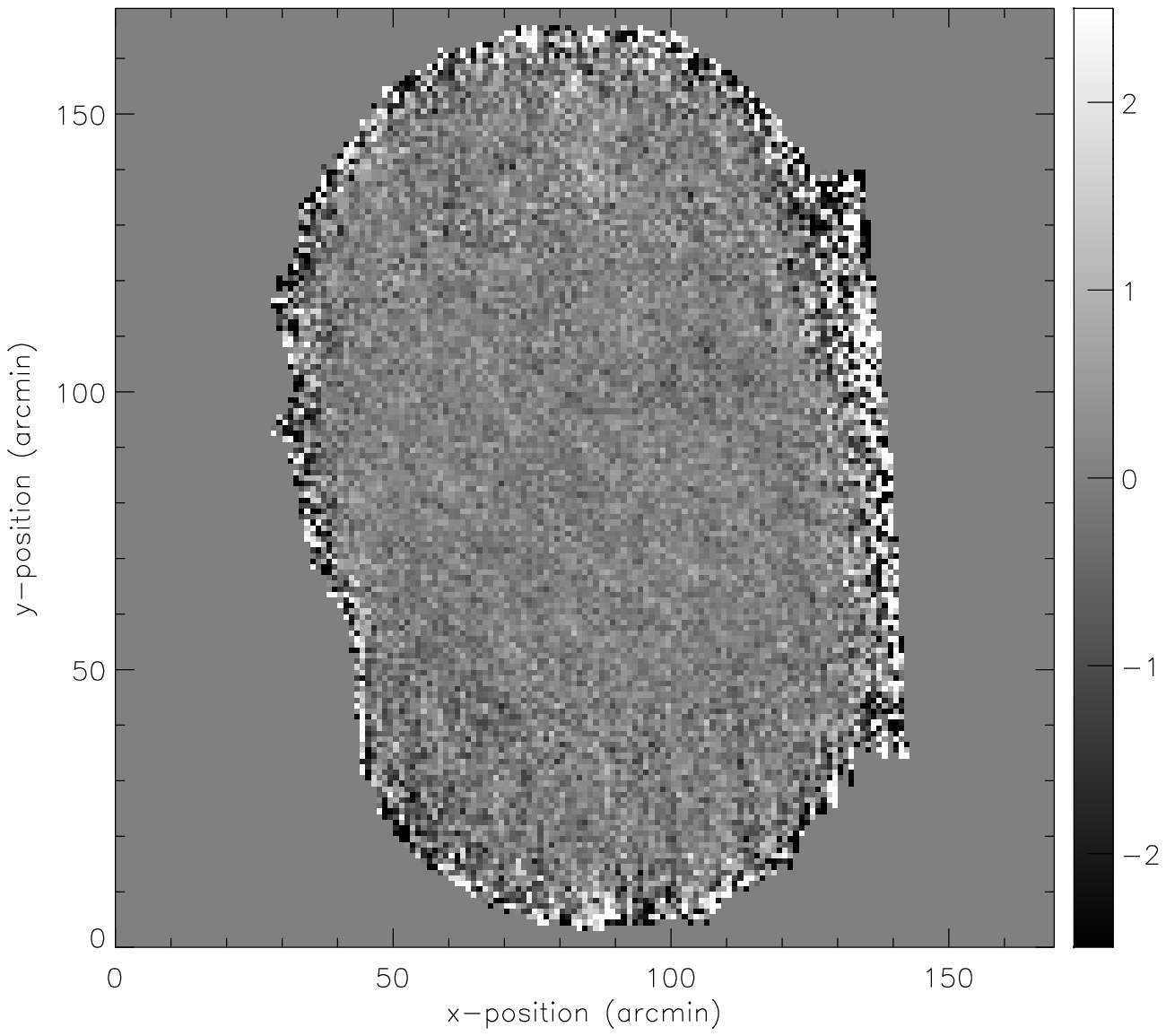}
      \includegraphics[scale=0.35]{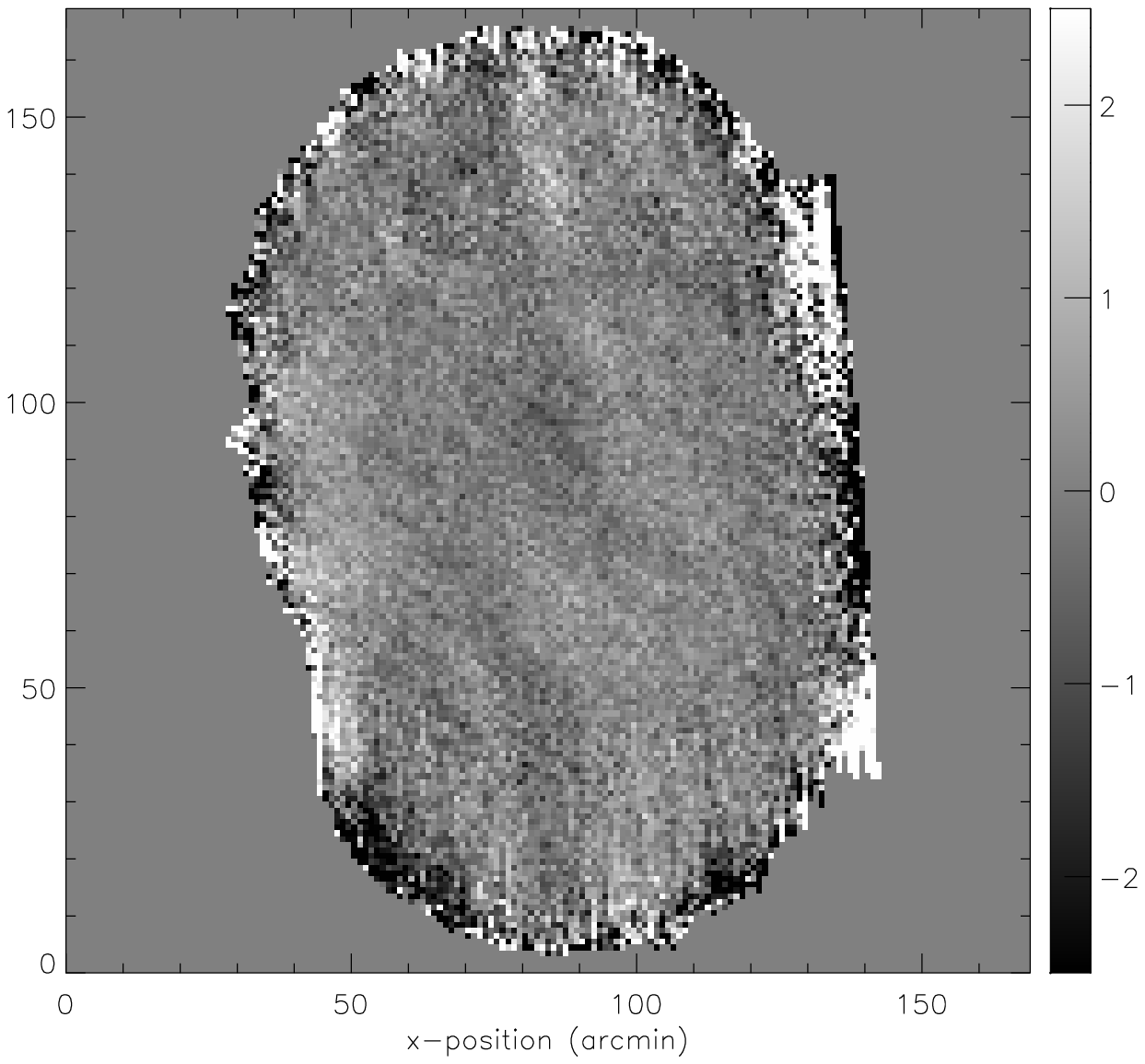}
      \includegraphics[scale=0.35]{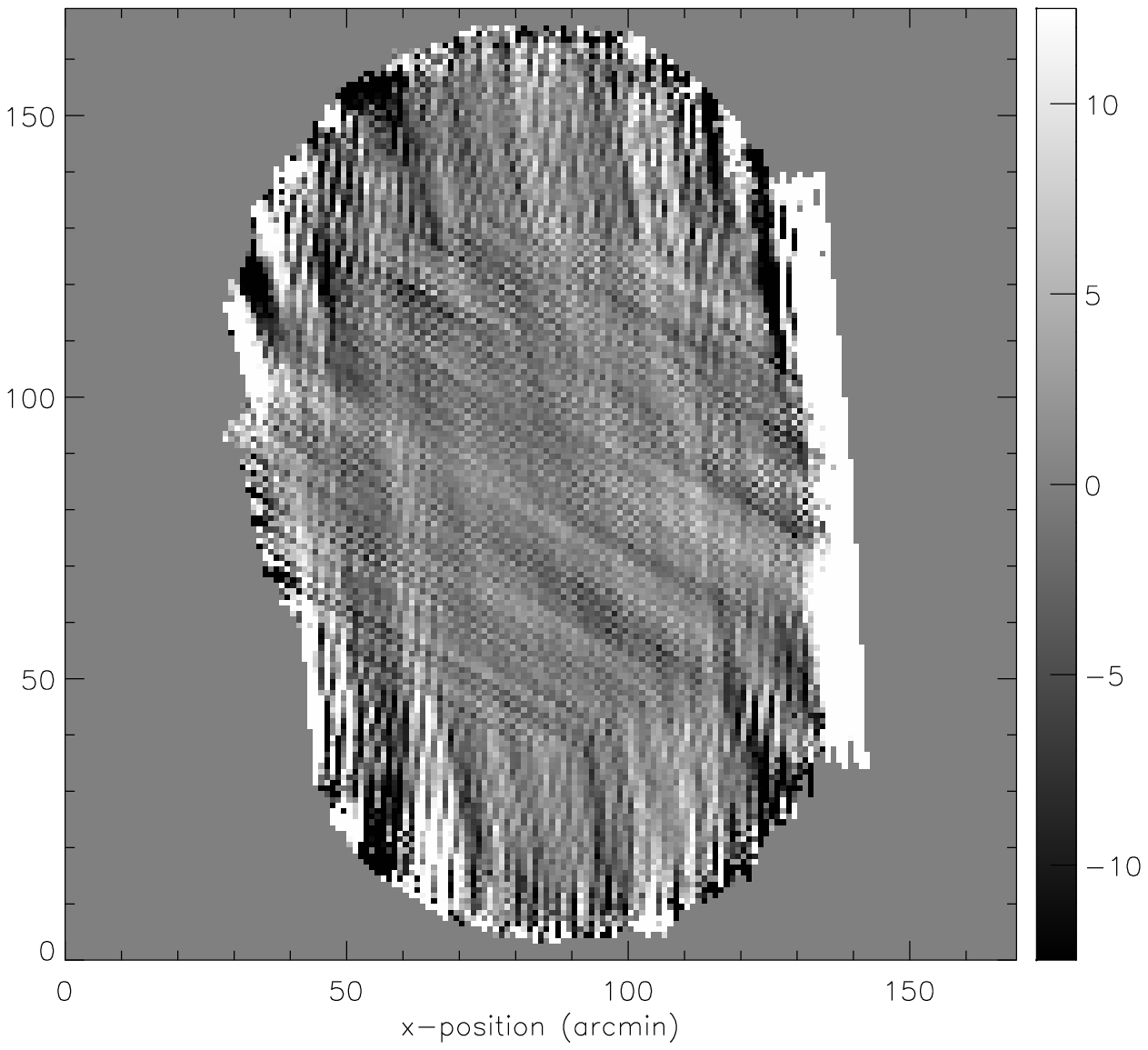}
    }
  \end{center}
  \caption{Noise-only simulations (like Figure~\ref{fig:mapnCasA_corr})
    for the BLAST05 \ivcg scanning
    configuration. The maps have a size of about 100$\arcmin$ across
    the short axis and 2.5$^\circ$ across the long axis. The pixel
    size here is 1$\arcmin$. The three panels show the map obtained
    with SANEPIC including noise correlations (left), SANEPIC with no
    noise correlations in the model (middle), and the simple co-added
    map (right).  Note the different brightness scale chosen for the last map
    due to its larger dynamic range.}
  \label{fig:mapnIVCG86}
\end{figure*}

Figure~\ref{fig:mapnIVCG86} shows the final noise maps in the three
cases for one realization of the simulations.
The simple re-projection map is obviously very stripy and would be of little
use as an estimate of the signal;
nevertheless it helps to visualize the directions of scanning in
the map. We can see two main directions covering the central
region of the map, oriented at about 50$^\circ$ to each other.  A third
scanning direction is also visible but has a much smaller weight. The
central part of the map is the cross-linked region, where we expect the
more optimal map-making procedures to excel.

In the map obtained using SANEPIC {\it without\/} considering noise
correlations between detectors (middle panel of Figure~\ref{fig:mapnIVCG86}),
some large scale noise is still visible in the
map, even if almost no residual striping is apparent in the cross-linked
region. Indeed, too much weight is given to the large timescales in
the timestreams (which are basically common between all the
detectors) as compared to the smaller timescales for which there are
more independent measurements, because the degree of correlation of
the noise is weaker. The residual noise at large angular scales is
much weaker in the SANEPIC map in which we account for the proper
correlations of the noise between detectors (left panel of
Figure~\ref{fig:mapnIVCG86}).  The noise in this map
looks particularly white.

\begin{figure}[t!]
  \begin{center}
    \includegraphics[width=\columnwidth]{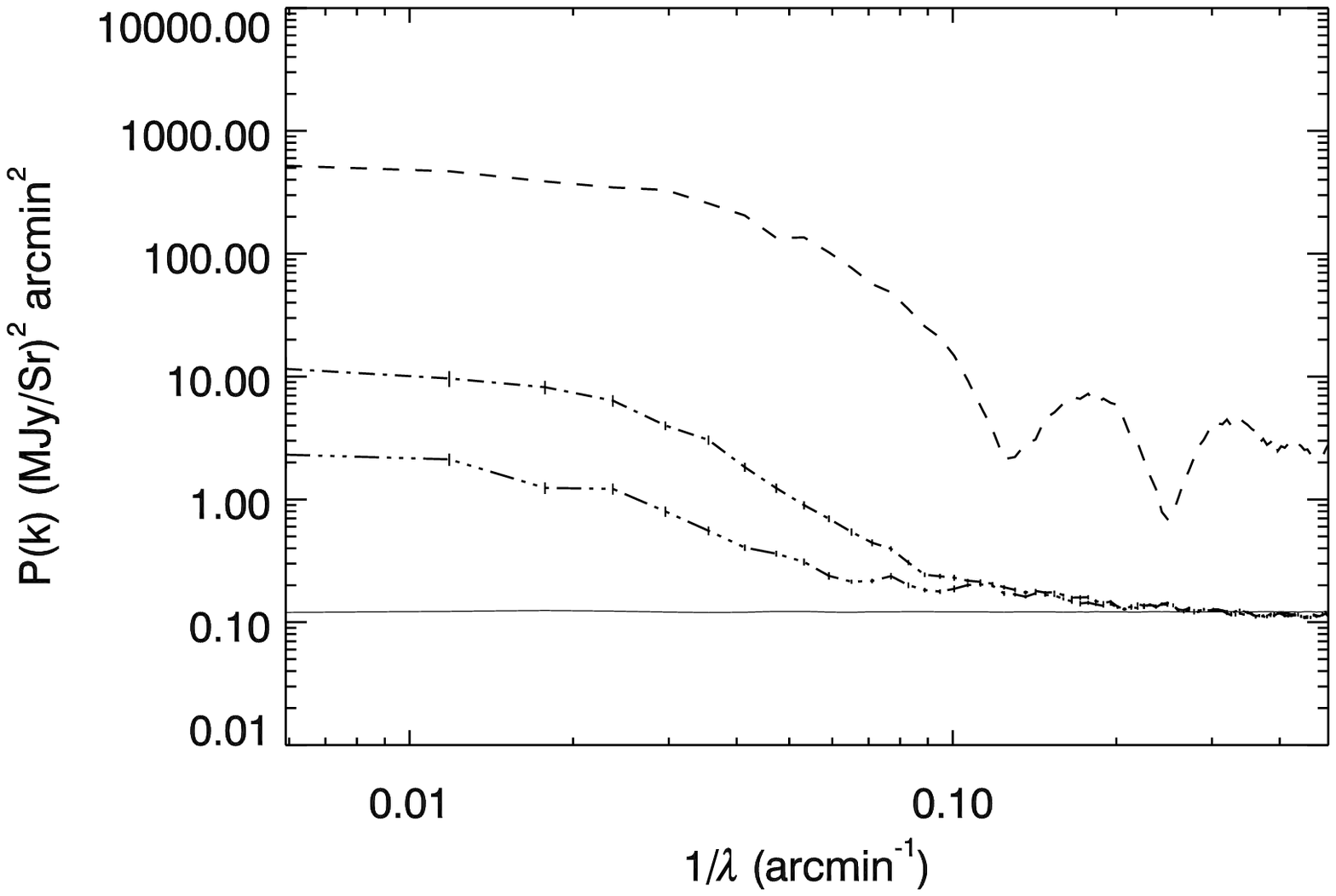}
  \end{center}
  \caption{Output power spectrum comparison of noise-only simulations
    after map-making (like Figure~\ref{fig:noise1dSpCasA}), for the
    BLAST05 \ivcg scanning configuration.  These power spectra are
    computed only in the cross-linked region of the maps, which
    forms a large disk of about 100$\arcmin$ diameter and can be easily
    identified in
    Figure~\ref{fig:mapnIVCG86}.  The dashed curve is for the simple
    re-projection map (right panel of Figure~\ref{fig:mapnIVCG86}),
    the dot-dashed curve is for SANEPIC {\it without\/} consideration of
    noise correlations between detectors
    (middle panel of Figure~\ref{fig:mapnIVCG86}),
    and the triple-dot-dashed curve is for SANEPIC {\it including\/} the correlation
    treatment (left panel of Figure~\ref{fig:mapnIVCG86}).
    The horizontal
    line indicates the level of white noise in the map predicted by
    the map-making procedure.  Error bars are estimated from the
    dispersion among measurements for all the realizations.  Residual low
    frequency noise in the optimal SANEPIC map is very low, thanks to
    the multiple scanning directions of this field. The situation is
    much better than for the \casa observational configuration, which
    had a single scan direction (Figure~\ref{fig:noise1dSpCasA}).}
  \label{fig:noise1dSpIVCG86}
\end{figure}

\begin{figure}[h!]
  \begin{center}
    \includegraphics[width=\columnwidth]{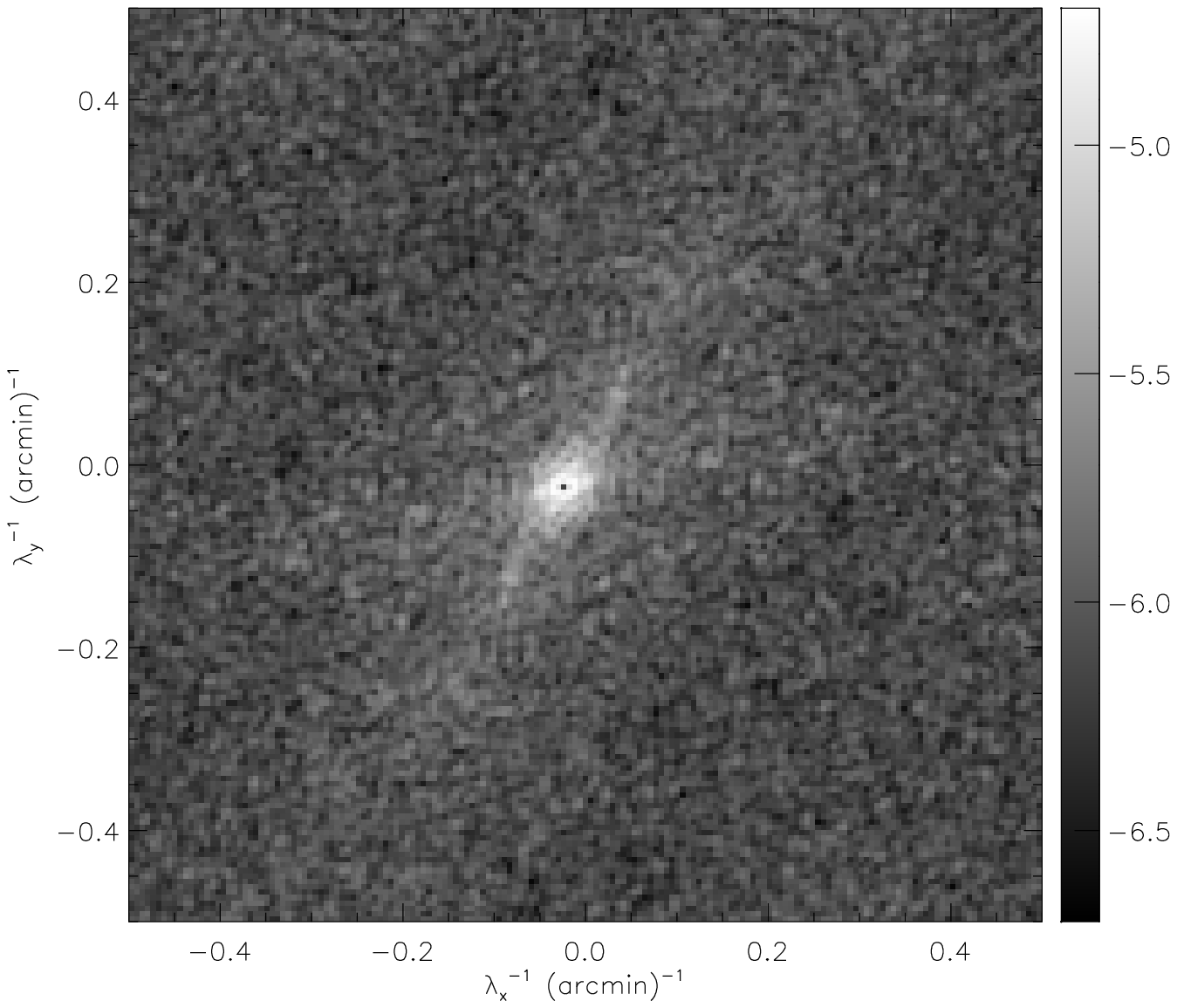}
  \end{center}
  \caption{2-D power spectrum of a noise-only simulation reduced using
    SANEPIC (like Figure~\ref{fig:Sp2dnoiseCasA_corr}) for the BLAST05
    \ivcg scanning configuration.}
  \label{fig:2dspectrumIVCG86}
\end{figure}

The noise level in each simulated map is quantified
in Figure~\ref{fig:noise1dSpIVCG86}, showing the 1-D power spectrum of
residual noise averaged over 20 simulations, and computed over the
cross-linked region (which has a diameter of about 100$\arcmin$).
While several orders of magnitude are gained in the noise power at all
scales using the SANEPIC ``no correlation'' method as compared to the simple
re-projection method, accounting for the correlations with SANEPIC
allows us to further reduce the noise power on scales ranging from 20$\arcmin$
to the size of the map by an additional factor of ${\sim}\,5$.  Toward smaller
scales, the gain between the SANEPIC correlation versus no correlation test
cases is still very significant down to about 10$\arcmin$, where the both
methods start to approach the white noise level. In the optimal map
obtained with SANEPIC the ratio between the noise power spectrum at
large scales and the white noise level is around 20, which is
relatively small. Figure~\ref{fig:2dspectrumIVCG86} shows the 2-D
power spectra of the noise in the SANEPIC map averaged over 20
realizations. As expected, the large scale noise is more important in
directions perpendicular to the scans.

\subsubsection{Signal-only timestreams}
\label{subsub:sigonlyIVCG}

\begin{figure*}[t!]
 \begin{center}\resizebox{16.5cm}{!}{
      \includegraphics[scale=0.35]{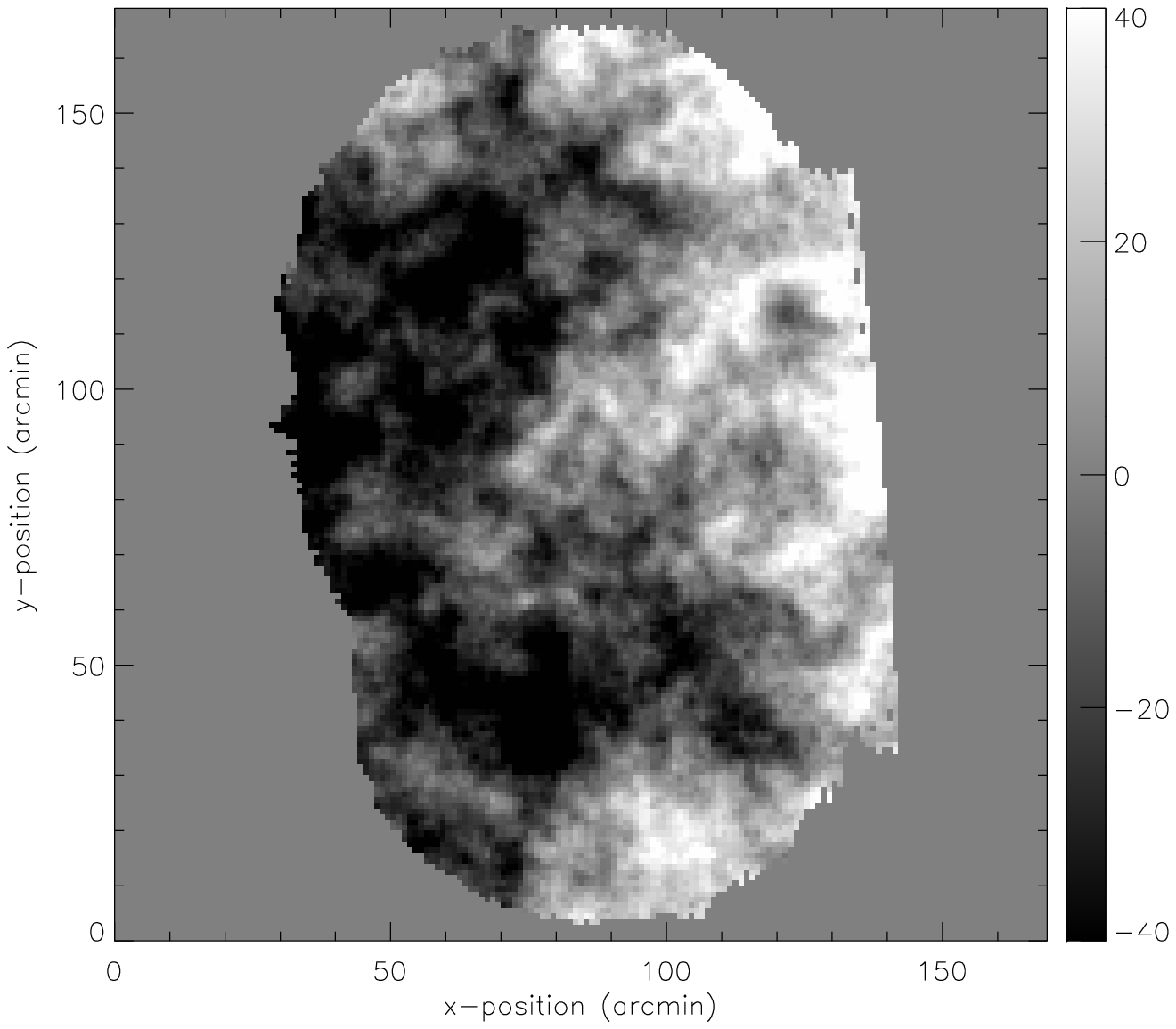}
      \includegraphics[scale=0.35]{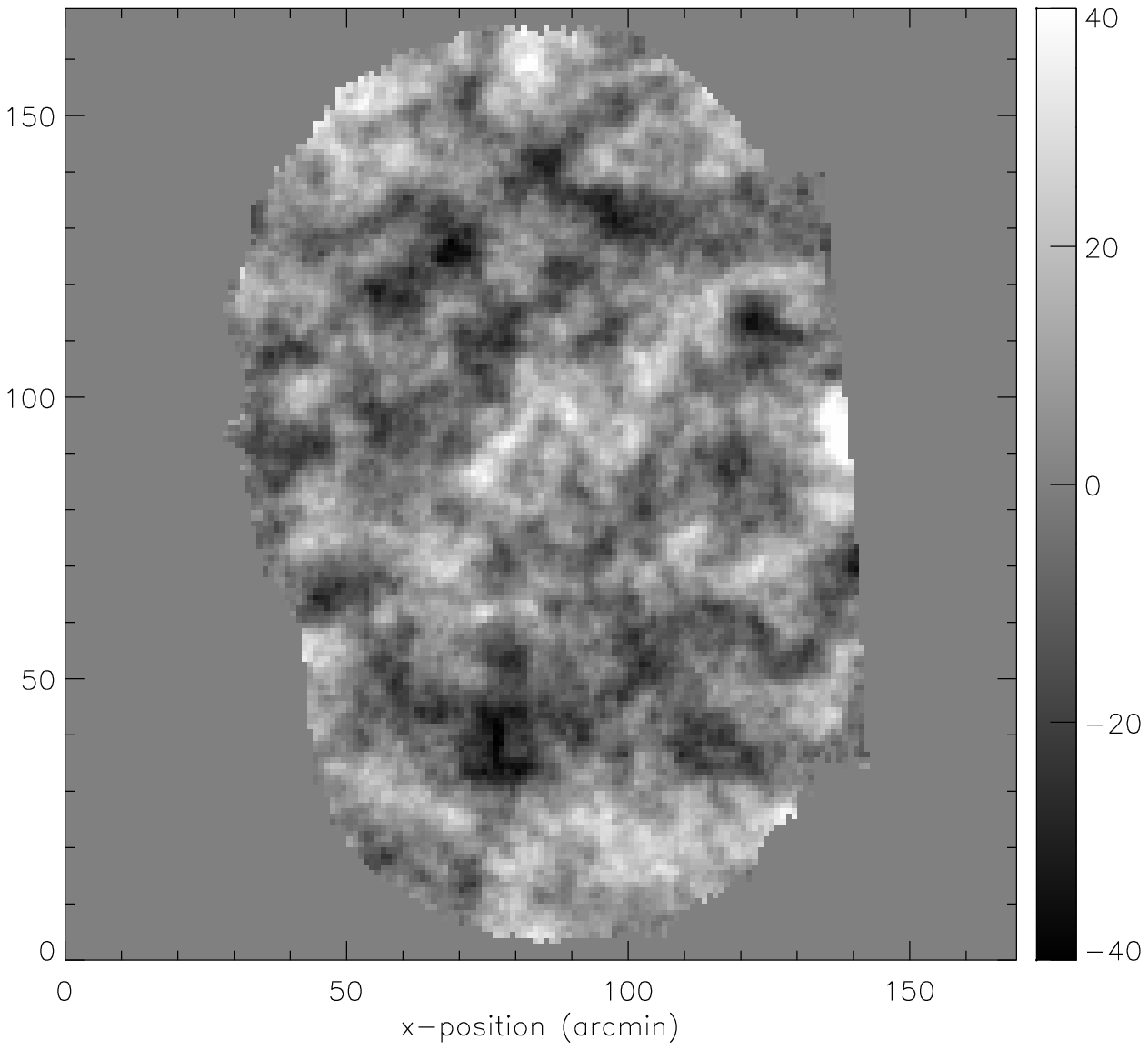}
      \includegraphics[scale=0.35]{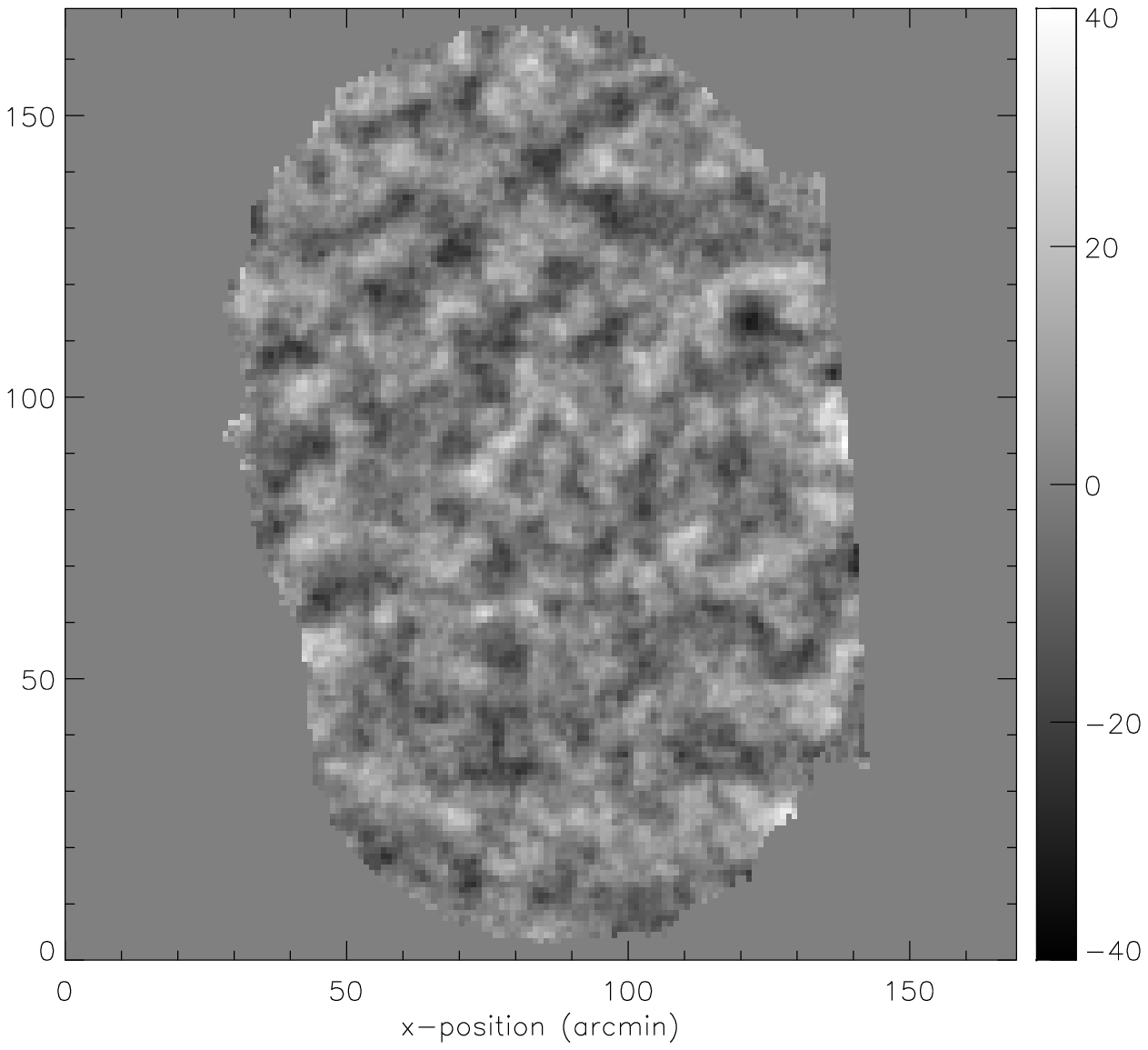}
    }
  \end{center}
  \caption{Signal-only simulations (like Figure~\ref{fig:mapsCasA})
    for the BLAST \ivcg configuration.
    The three panels show the input signal map (left), full
    SANEPIC (middle) and simple common-mode subtraction followed by
    application of SANEPIC without correlations (right).}
  \label{fig:mapsIVCG86}
\end{figure*}

We now focus on signal-only simulations for G86.  As in the case of the
\casa configuration, we compare the performance of SANEPIC with
respect to the simpler common-mode subtraction method.
Figure~\ref{fig:mapsIVCG86} shows the pure signal input map for one
realization of the simulations (left panel) compared with two recovered maps.
The first output map is obtained with SANEPIC
including correlations between detectors (middle panel), while the second is
obtained by subtracting the common mode between all detectors, followed by
applying SANEPIC, but neglecting noise correlations between detectors (right
panel).  We see that the very
largest scales are not recovered by SANEPIC.  This is because of the
weak filtering applied to the timestreams and, to a greater extent,
because of inversion problems with SANEPIC on scales of the order of
the size of the map (these scales are very poorly constrained by the
map-making procedure). In the common-mode subtraction method, only the
very largest scales are suppressed by the map-making procedure itself,
but the filtering effect is more dramatic and extends to much smaller
scales.

\begin{figure}[h!]
  \begin{center}
    \includegraphics[width=\columnwidth]{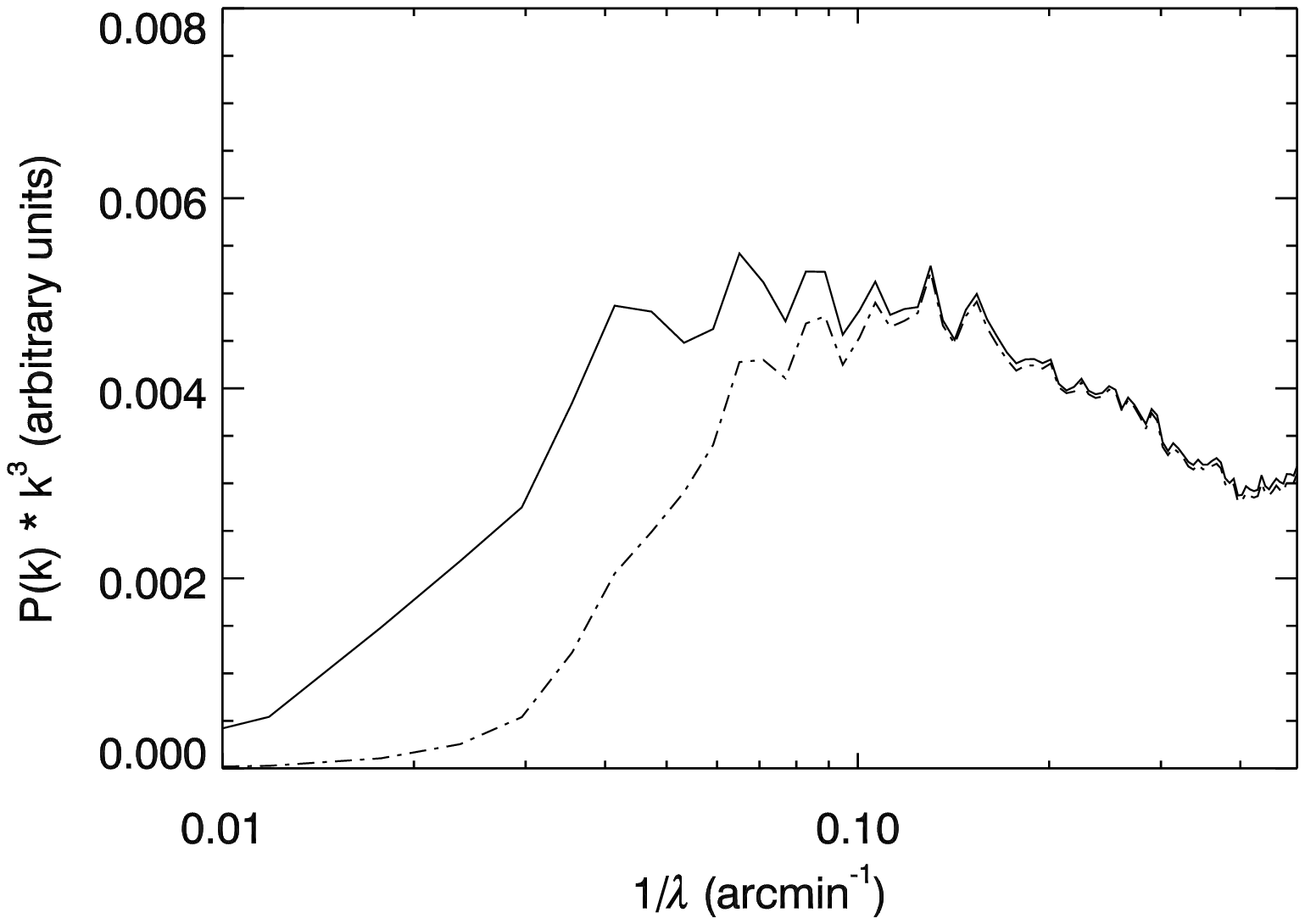}
  \end{center}
  \caption{Power spectrum comparison of the SANEPIC map (solid) versus the
    simple common-mode removal map (dot-dashed) for signal-only simulated
    data, as in Figure~\ref{fig:1DpowerCasA}, but for the BLAST05 \ivcg
    scanning configuration.  Power spectra are multiplied by $k^3$, so that
    a flat line would indicate no filtering.  The drop off at small angular
    scales is due to the BLAST05 PSF.}
  \label{fig:1DpowerIVCG86}
\end{figure}

The effective filtering is quantified in
Figure~\ref{fig:1DpowerIVCG86}, which shows the power spectra of output
maps averaged over 20 simulations of pure signal timestreams. Again the
power spectra are multiplied by $k^3$ for comparison purposes.
In this case SANEPIC works well on scales up to about half a degree, above
which it fails to recover structure in the map;
this limit corresponds to scales of about a
quarter of the map and larger.  The filtering effect is much more
pronounced in the map produced with the common-mode subtraction method,
being strong for all scales above around
14$\arcmin$.  This is of course expected, since subtracting the average of the
array strongly reduces the signal on scales larger than the array
size. Therefore, in order to recover large and intermediate scale
structure in the maps, it is beneficial to use SANEPIC instead of
other methods that are based on simply filtering the data.

\subsection{Advantages of cross-linking}

The relative level of residual noise at low spatial frequency in the
maps is significantly reduced in the \ivcg observational configuration
as compared to the \casa configuration. The fundamental difference is
that the \ivcg observations contain multiple (essentially two for most of the
data) scanning directions, while the \casa observations are realized
with only two passes across the field in the same direction.  Multiple
scanning directions give a huge number of additional constraints for
the map-making procedure.  In particular, large scale structures in the
map are much better recovered in directions parallel to scans, because
the noise there is smaller.  Thus having multiple scanning angles allows for
recovery of the sky fluctuations for all directions, and ends up giving
almost no weight to the individual loosely constrained cross-scan k-modes.

Differences in the results for maps from these two example scanning strategies
can be quantified in two ways. First of all, for the \ivcg scanning
strategy, the transition between white noise and ``excess'' large
scale noise in the map occurs at a scale around 10$\arcmin$, while the
same transition occurs at a scale of around 3$\arcmin$ for the \casa
scanning strategy.  Secondly, the ratio between large scale noise power
and white noise power is larger by more than two orders of magnitude
for \casa than for G86.  On the other hand, some caution should be
taken to not over-interpret this comparison, because the pixel size we used
is larger for the \ivcg map (1$\arcmin$) as compared to the \casa map
(25$\arcsec$), and therefore the number of crossings per pixel is
greater for the \ivcg map.  Nevertheless, this simulation exercise has
demonstrated that cross-linking in the map is extremely beneficial,
particularly for recovering the large scale structures in the map.

\subsection{Map-making transfer function}

When carrying out a complex data processing procedure, it is important
to check whether the results are biased in any way.  We have found
that the transfer function of the map-making procedure, defined as the
ratio between the amplitude of fluctuations in the output pure signal
map relative to the input map, is not always exactly unity, even for
intermediate and small angular scales in the map.  For example, in the \casa
configuration at 250$\,\mu$m the fluctuation amplitudes in the final
map are reduced by 3\% on average as compared to the input map, almost
uniformly across spatial frequencies and directions. This global
discrepancy reaches the level of 9\% at 500$\,\mu$m.  Moreover, it is also
present for the \ivcg configuration.  We believe that this reduction
is due to the fact
that the pixel-pixel covariance is ill-conditioned and numerical
imprecision occurs in the matrix inversion.  We find that the bias
tends to be smaller when the number of detectors is larger and also
when the number of constraints increases, like when we have multiple
scanning directions, or when we map isolated bright sources (presented
in Truch et al.~2007) and constrain the data outside a defined region
to have a constant flux (see Section~\ref{sub:pixconstr} for details
of this procedure).  Since this bias can be estimated using
simulations, it is straightforward to correct for.  We have found that
it is not always important, e.g.~for the large Galactic map in the Vulpecula
region analyzed in Chapin et al.~(2007) the bias is negligible.

\section{Application to real BLAST05 data}
\label{sec:AppliData}

Now that we have looked at signal-only and noise-only simulations, we
now turn to real BLAST data.
In this section, we show maps of two example fields from the BLAST05
data which have been obtained using SANEPIC.

\subsection{Cassiopeia A}
\label{sub:resCasA}

\begin{figure}[t!]
  \begin{center}
    \includegraphics[width=\columnwidth]{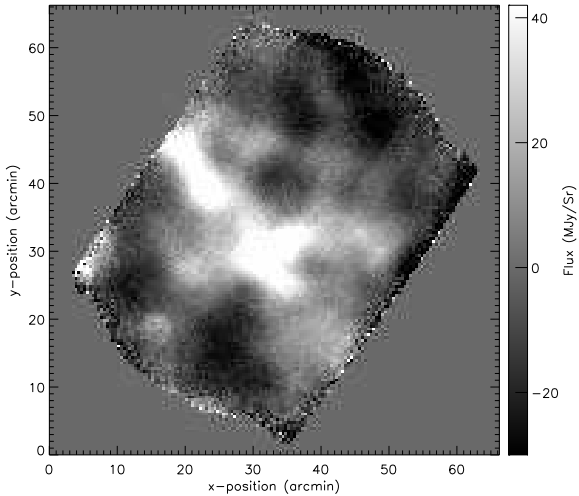}
  \end{center}
  \caption{Map of the \casa supernova remnant at 250$\,\mu$m made from BLAST05
    data using SANEPIC including noise correlations between
    detectors.  The map is represented in Galactic coordinates with
    25$\arcsec$ pixels and has a size of about $0.5\,{\rm deg}^2$.}
  \label{fig:mapCasA250}
\end{figure}

Figure~\ref{fig:mapCasA250} shows the map obtained from the
observations of the \casa field at 250$\,\mu$m using SANEPIC including full
consideration of the noise correlated between detectors.  Detailed analysis
of the maps at the three wavelengths is described in Hargrave
et al.~(in preparation).
The properties of the noise and the transfer function of the signal in the
map have been studied in detail from Monte Carlo simulations in Section
\ref{sub:CasACase}. Results from such simulations are used to characterize
the map.  The power spectrum of this map has been compared to results
from simulations in Figure~\ref{fig:noise1dSpCasA}.
The map structures are relatively smooth, due to the BLAST05 point spread
function, which has a width of the order of 3$\arcmin$, causing the
drop in the 1-D power spectrum below those spatial scales.  The signal
clearly dominates over noise on angular scales larger than about
3$\arcmin$, and the diffuse structure should be reliable up to a large
fraction of the overall map size.

\subsection{Vulpecula region}
\label{sub:Vulpecula}

Another field observed during the BLAST05 campaign is centered in
the Galactic Plane close to the open cluster NGC 6823 in the
constellation of Vulpecula. The region mapped has a size of about
$4\,{\rm deg}^2$ and was chosen for its high-mass star formation
activity.  Complete analysis of this observed field is presented in
\cite{chapin}.

A few hours of these data were taken at different time intervals
during the flight. By design, this field has been observed with very
different scanning directions, and is therefore it should be possible to
recover diffuse large scale structures.

\begin{figure*}[t!]
  \begin{center}\resizebox{16.5cm}{!}{
      \includegraphics[scale=0.35]{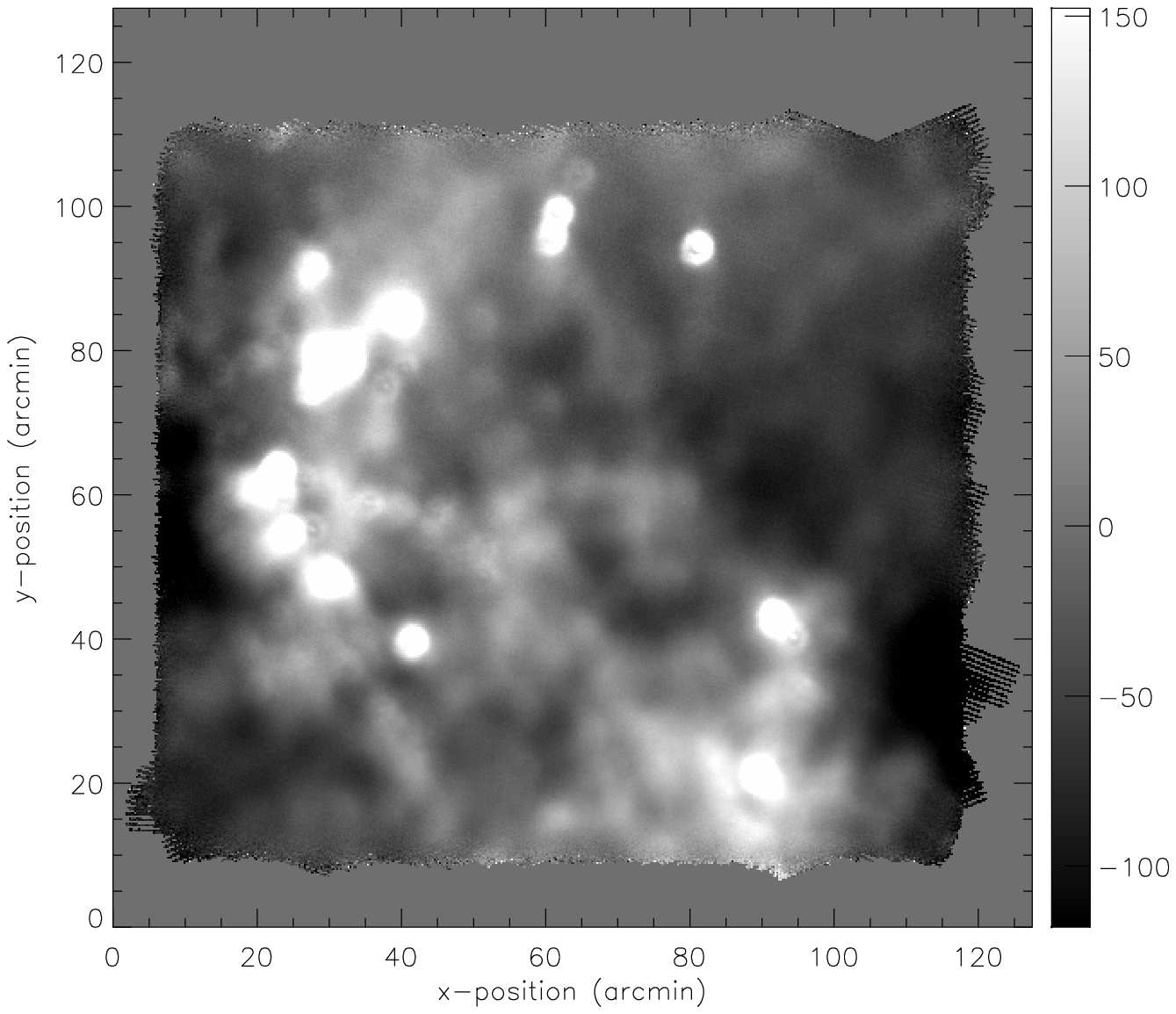}
      \includegraphics[scale=0.35]{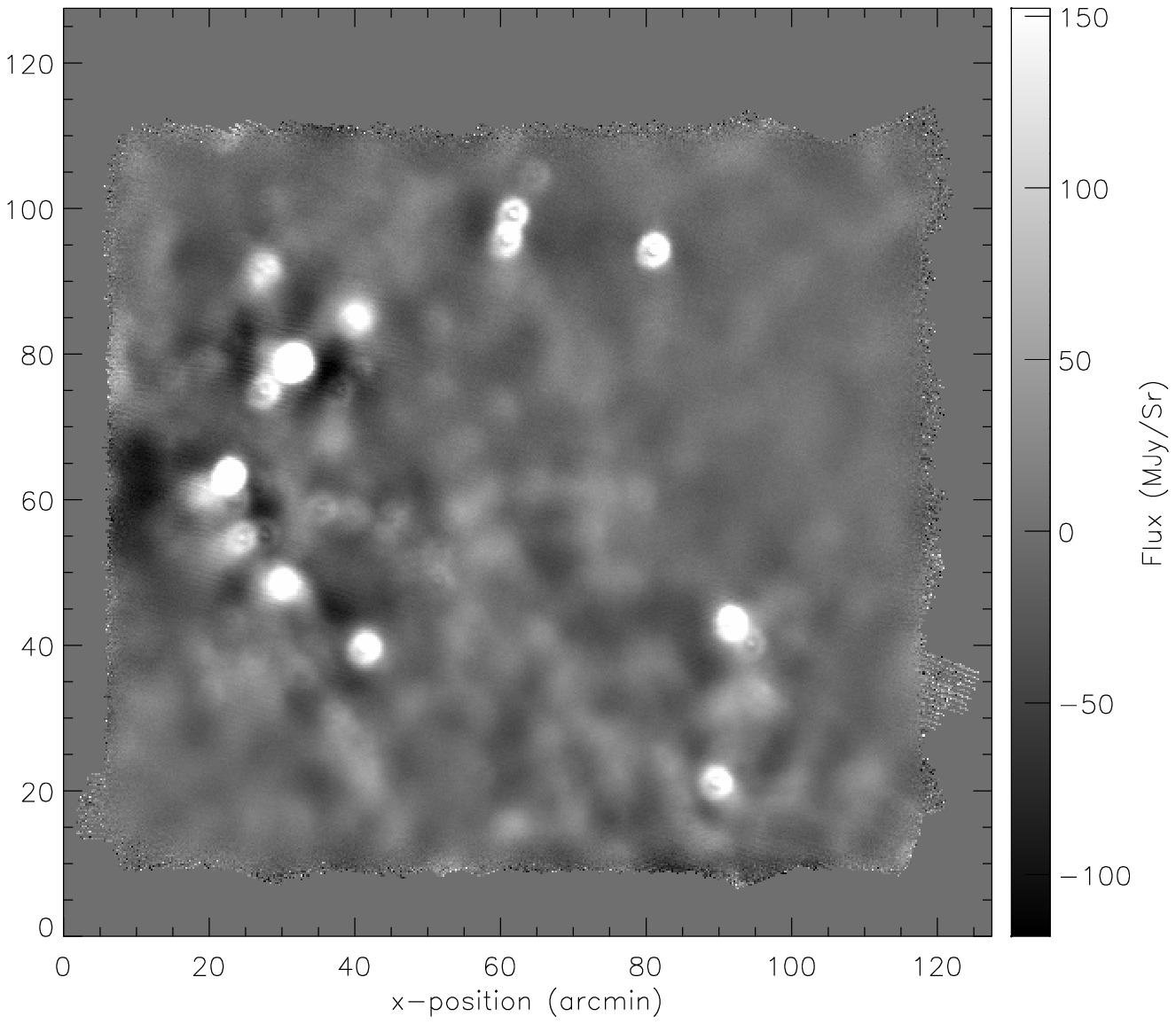}
    }
  \end{center}
  \caption{Maps of a Galactic Plane region near NGC 6823 in the
    Vulpecula constellation derived from BLAST05 observations at
    250$\,\mu$m. The two maps are obtained using two different
    methods: SANEPIC (left panel); and simple reprojection after
    removing the array average signal at each timestep from all the
    timestreams, together with filtering (right panel).  Maps are presented in
    equatorial coordinates with 15$\arcsec$ pixels. The region mapped
    has a size of about $4\,{\rm deg}^2$.  Numerous point sources have been
    identified in the field (see Chapin et al.~2007), their tell-tale
    shape in the map resulting from the PSF of the
    BLAST05 optics (see Truch et al.~2007).}
  \label{fig:mapL5B250}
\end{figure*}

The map of the observed region at 250$\,\mu$m obtained with SANEPIC is
shown in Figure~\ref{fig:mapL5B250} (left panel).  For comparison
(right panel), we have computed another map using a much simpler method
which consists of removing the array average signal at each timestep
for all of the timestreams and reprojecting the data onto the map
after filtering.  This is like the ``sky removal'' procedure often
carried out for ground-based submillimeter data (see also
Section~\ref{sec:CasAsignal}).
One can see that it
suppresses almost all the diffuse structure in the map.

No residual striping is visible in the map obtained with SANEPIC (left
panel of Figure~\ref{fig:mapL5B250}),
mainly due to the presence of multiple scanning directions for this
field.  Large scale structures in the map are successfully recovered
with SANEPIC, as can easily be seen by comparing with the right panel
of Figure~\ref{fig:mapL5B250}.  This recovery applies to scales which are
significantly larger than the array size.
However, the resulting effective filtering after applying the
``array average subtraction'' method induces negative signals near
bright sources in the map, while no such filtering effect is seen in the
SANEPIC map (except perhaps near the edges of the map). This shows
that optimal map-making methods (in the sense of least squares) like
SANEPIC are better suited to recover point sources in the maps as well as
diffuse structures.

\section{Conclusions}

Large format detector arrays operating at far-IR and submillimeter
wavelengths are becoming the norm, rather than the exception.
Ground-based instruments are plagued with common-mode emission arising
from the Earth's atmosphere.  And as we have found with BLAST, the
same applies to high frequency balloon-borne instruments, where we see
correlated noise from thermal as well as atmospheric effects.  There
is an expectation that even upcoming satellites might be faced with
similar issues, because of thermal variations in the spacecraft, for
example.  In general, we expect that correlated noise between
detectors will be a major issue which all such experiments have to
deal with, and we expect that the SANEPIC approach, which we have
described here, will be widely applicable.  Indeed, there is evidence
from existing arrays (e.g.,~SHARC-II and AzTEC) that once there are
many detectors, there are {\it multiple\/} correlations between sub-sets of
the detectors, as well as an overall common-mode term.  Consequently, one sees
correlations between contiguous blocks of detectors on the array, or
sets of detectors which share amplifiers or are otherwise coupled
through the electronics.  Provided that these correlations can be
investigated and their behavior modelled, it is straightforward to
extend the SANEPIC approach to deal with several distinct sources of
correlated noise.  Hence we expect the SANEPIC approach to be applicable to
future instruments such as SCUBA-2, SPIRE, ACT, {\sl Planck\/} HFI and
others.  There are also many experiments being planned which use large
detector arrays to perform sensitive polarization measurements, and we
see no reason why the SANEPIC approach could not also be
extended to polarimetry.

\acknowledgments

The BLAST collaboration acknowledges the support of NASA through grant
numbers NAG5-12785, NAG5-13301 and NNGO-6GI11G, the Canadian Space
Agency (CSA), Canada's Natural Sciences and Engineering Research
Council (NSERC), and the UK Particle Physics \& Astronomy Research Council
(PPARC).  We would also like to thank the Columbia Scientific Balloon
Facility (CSBF) staff for their outstanding work.
We are grateful to Matthew
Hasselfield and Duncan Hanson for valuable discussions which contributed
to the development of SANEPIC.  LO acknowledges
partial support by the Puerto Rico Space Grant Consortium and by the
Fondo Istitucional para la Investigacion of the University of Puerto
Rico.  CBN acknowledges support from the Canadian Institute for
Advanced Research.  This research made use of Westgrid computing
resources, which are funded in part by the Canada Foundation for Innovation,
Alberta Innovation and Science, BC Advanced Education, and the participating
research institutions.

\bibliographystyle{apj}

\end{document}